\documentclass[]{aa}  

\usepackage{subfig}
\usepackage{xcolor}
\usepackage{graphicx}
\usepackage{txfonts}
\usepackage{natbib,twoopt}
\usepackage[breaklinks=true]{hyperref} 
\bibpunct{(}{)}{;}{a}{}{,}             
\newcommand{\deltaL}{$\log \Delta L_{500}$}
	
\makeatletter
  \newcommandtwoopt{\citeads}[3][][]{\href{http://adsabs.harvard.edu/abs/#3}%
    {\def\hyper@linkstart##1##2{}%
     \let\hyper@linkend\@empty\citealp[#1][#2]{#3}}}
  \newcommandtwoopt{\citepads}[3][][]{\href{http://adsabs.harvard.edu/abs/#3}%
    {\def\hyper@linkstart##1##2{}%
     \let\hyper@linkend\@empty\citep[#1][#2]{#3}}}
  \newcommandtwoopt{\citetads}[3][][]{\href{http://adsabs.harvard.edu/abs/#3}%
    {\def\hyper@linkstart##1##2{}%
     \let\hyper@linkend\@empty\citet[#1][#2]{#3}}}
  \newcommandtwoopt{\citeyearads}[3][][]%
    {\href{http://adsabs.harvard.edu/abs/#3}
    {\def\hyper@linkstart##1##2{}%
     \let\hyper@linkend\@empty\citeyear[#1][#2]{#3}}}

\makeatother

\begin{document}

   \title{The impact of assembly history on the X-ray detectability of halos}

    \subtitle{From galaxy groups to galaxy clusters}
   \author{I. Marini\thanks{Corresponding author: \texttt{ilaria.marini@eso.org}}
          \inst{1}
          \and
          P.~Popesso\inst{1, 2}
          \and
          K. Dolag\inst{3, 4, 2}
          \and
          V. Biffi\inst{5, 6}
          \and
           S. Vladutescu-Zopp\inst{3}
          \and
          T.~Castro\inst{5, 6, 7, 8}
          \and
          V. Toptun\inst{1}
          \and
          N. de Isídio\inst{1}
          \and 
          A. Dev\inst{11}
          \and
          D. Mazengo\inst{1, 12}
          \and
          J. Comparat\inst{13}
          \and
          C. Gouin\inst{9, 10}
          \and
          N. Malavasi\inst{13}
          \and
          A. Merloni\inst{13}
          \and
          T. Mroczkowski\inst{1}      
          \and\\
          G. Ponti\inst{14, 13, 15}
          \and
          S. Shreeram\inst{13} 
          \and
          Y. Zhang\inst{13}
          }

   \institute{European Southern Observatory, Karl Schwarzschildstrasse 2, 85748, Garching bei M\"unchen, Germany\\
                \email{ilaria.marini@eso.org}
         \and
            Excellence Cluster ORIGINS, Boltzmannstr. 2, D-85748 Garching bei M\"unchen, Germany
        \and
            Universitäts-Sternwarte, Fakultät für Physik, Ludwig-Maximilians-Universität München, Scheinerstr.1, 81679 München, Germany
        \and 
            Max-Planck-Institut für Astrophysik, Karl-Schwarzschildstr. 1, 85741 Garching bei M\"unchen, Germany
        \and
            INAF – Osservatorio Astronomico di Trieste, Via Tiepolo 11, 34143 Trieste, Italy
        \and 
            IFPU – Institute for Fundamental Physics of the Universe, Via Beirut 2, I-34014 Trieste, Italy
        \and
            INFN, Sezione di Trieste, Via Valerio 2, 34127 Trieste TS, Italy        
        \and
            ICSC - Centro Nazionale di Ricerca in High Performance Computing, Big Data e Quantum Computing, Via Magnanelli 2, Bologna, Italy
        \and
            Sorbonne Université, UMR7095, Institut d’Astrophysique de Paris, 98 bis Boulevard Arago, F-75014 Paris, France
        \and
            Université Paris-Saclay, CNRS, Institut d’Astrophysique Spatiale, 91405 Orsay, France
        \and
            International Centre for Radio Astronomy Research, University of Western Australia, M468, 35 Stirling Highway, Perth, WA 6009, Australia
        \and
             Department of Physics, College of Natural and Mathematical Sciences, University of Dodoma, P.O. Box 338, Dodoma, Tanzania
        \and
            Max Planck Institute for Extraterrestrial Physics, Giessenbachstrasse 1, 85748 Garching, Germany
        \and
            INAF – Osservatorio Astronomico di Brera, Via E. Bianchi 46, 23807 Merate (LC), Italy
        \and
            Como Lake Center for Astrophysics (CLAP), DiSAT, Università degli Studi dell’Insubria, via Valleggio 11, I-22100 Como, Italy
             }

   \date{Received 21 March 2025 ; accepted 17 April 2025 }
 
  \abstract
    {Galaxy groups represent a significant fraction of the halo population, playing a crucial role in galaxy formation and evolution. However, their detection in X-rays remains challenging, raising questions about the physical mechanisms driving their detectability in current surveys. Using the Magneticum simulations, we construct a mock X-ray lightcone of the local Universe ($z<0.2$) to investigate the selection function of galaxy groups and clusters. We find that the central supermassive black holes (SMBH) activity is a key driver of baryon depletion, but late-time mergers boost X-ray brightness by replenishing the gas reservoir in the halos, highlighting the interplay between feedback processes and the environment. Our analysis shows that X-ray bright groups experience sustained late-time mass accretion, maintaining higher gas fractions and fuelling the central SMBH, further increasing the X-ray emissivity in the core. In contrast, X-ray faint groups form earlier and lose most of their gas over time, resembling fossil groups. These faint groups reside preferentially in under-dense regions, unable to re-accrete efficiently baryons at later times. Magneticum predicts strong anti-correlations between gas fraction (or X-ray luminosity) and SMBH mass, stellar mass (both in the central galaxy and intracluster light), and group richness at fixed halo mass. We derive predictions on the hot gas fraction at fixed halos mass (e.g. a group of total mass $M_{500}=10^{13} M_{\odot}$ can have hot gas fractions in the range $f_\mathrm{gas}=0.02-0.06$ and a central SMBH with a median mass of $M_\mathrm{BH}=10^9 M_{\odot}$ and a scatter of $0.5$ dex) compatible with the most recent measurements of the baryonic fraction. These findings will aid the interpretation of future X-ray surveys, demonstrating the power of simulation-based inference.
        }


   \maketitle
%
\begin{figure*}
    \centering    
    \subfloat{\includegraphics[width=\textwidth,trim={0 13.cm 1cm  0},clip]{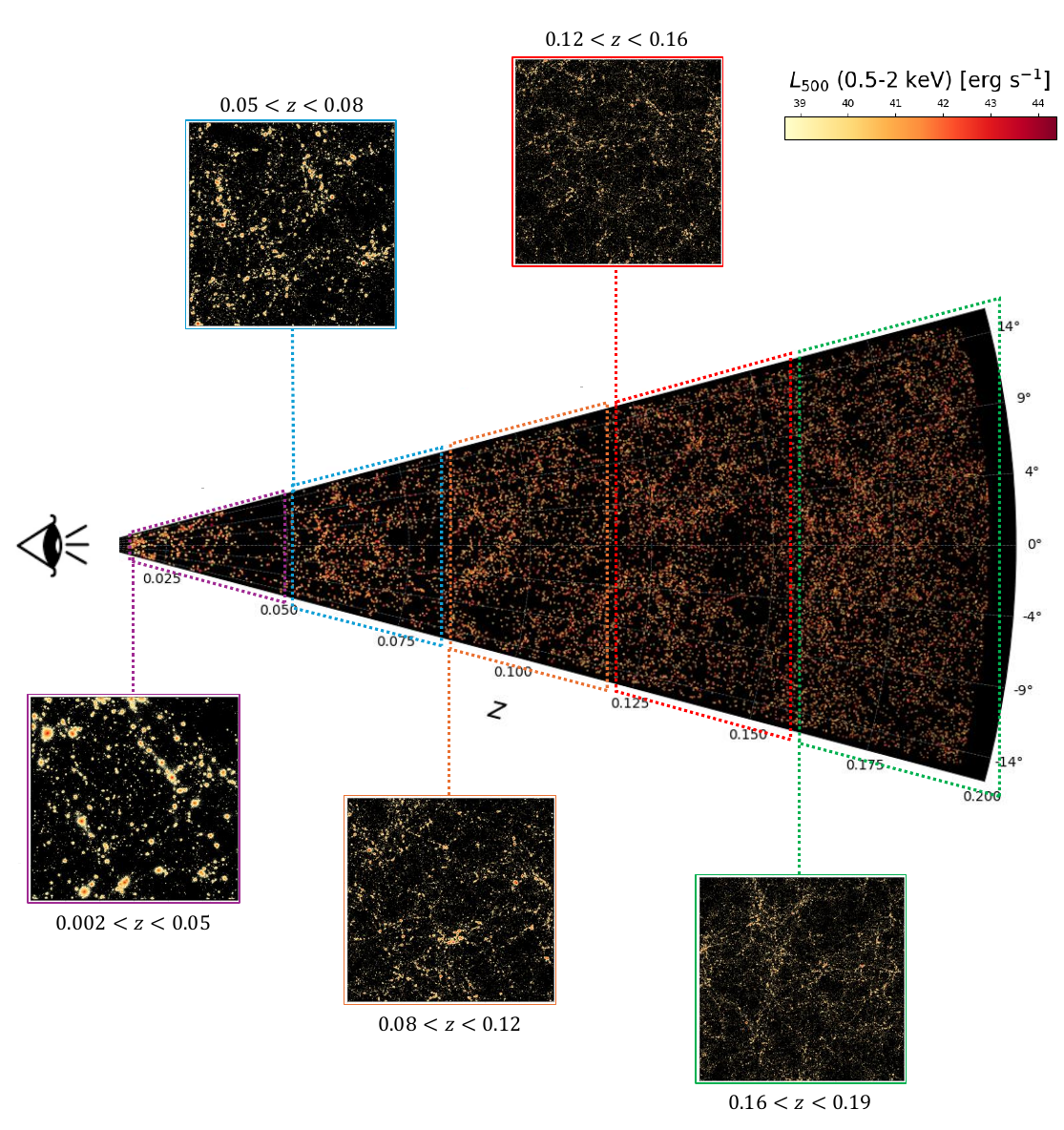}} 
    \caption{Schematic representation of the X-ray lightcone LC30, where we show the distribution of galaxy groups and clusters with $M_{200} > 10^{12.5} M_{\odot}$ and colour-coded for their intrinsic X-ray luminosity in $0.5-2.0$ keV within $R_{500}$. The X-ray emission is based on PHOX's photon list. Each smaller panel represents the projected X-ray emission corresponding to the shells at different snapshots forming LC30 and has an angular scale of $30\times 30$ square degrees. }
    \label{fig:LC30_xray}
\end{figure*}
\section{Introduction}
Galaxy groups are fundamental building blocks of the cosmic web, bridging the gap between isolated galaxies and massive clusters \citep[see a comprehensive review][]{lovisari_physical_2021} and serve as ideal laboratories for studying galaxy evolution, environmental effects, and the role of feedback mechanisms. They are gravitationally bound structures of a few to dozens of galaxies, typically with total masses ranging from $10^{13}M_{\odot}$ to $10^{14} M_\odot$ \citep{tully_nearby_1987}. Unlike clusters, where hot plasma dominates the intracluster medium (ICM) in a relatively stable hydrostatic state \citep{borgani_thermodynamical_2008}, groups exhibit a more diverse range of properties in the intragroup medium \citep[IGrM;][]{zuhone_effects_2023, lovisari_physical_2021}. Understanding the physical processes governing their evolution is crucial for interpreting observations from current and future surveys.  
\par
One of the primary ways to study galaxy groups is through their X-ray emission, which originates from hot, diffuse gas in the IGrM. Moreover, at low temperatures, line cooling becomes very important, and the emissivity (both in soft and bolometric bands) becomes strongly metallicity- and temperature-dependent \citep[see Fig. 1 in][]{lovisari_scaling_2021}. In this context, X-ray detectability refers to the ability to identify galaxy groups through their extended X-ray emission, which depends on factors such as gas density, temperature, metallicity, and the sensitivity of current and future X-ray observatories. Observations from missions such as Chandra, XMM-Newton, and eROSITA have revealed significant variations in the X-ray properties of groups, with some systems appearing bright and extended \citep{sun_chandra_2009, andreon_variegate_2017, bulbul_srgerosita_2024, bahar_srgerosita_2024}. In contrast, many other galaxy groups remain faint or undetected \citep{zheng_measuring_2023, popesso_x-ray_2024, popesso_average_2024, marini_detecting_2024, Marini2025_erratum, li_robust_2024}. While an optical follow-up to detect them is possible \citep{bulbul_erosita_2022,ota_erosita_2023, balzer_first_2025}, a considerable theoretical effort is in place to understand such systems \citep[e.g.][]{oppenheimer_eagle_2020, ragagnin_simulation_2022,  cui_hyenas_2024, hough_simba-c_2024, bigwood_case_2025}. These differences are thought to be due to the structural properties of groups, driven by non-gravitational processes that alter the thermodynamic state of the hot gas and the overall baryon content in groups and clusters \citep{wu_non-gravitational_2000, borgani_effect_2002, davies_gas_2019, oppenheimer_eagle_2020}. In principle, their assembly history, merger activity, and impact of feedback from active galactic nuclei (AGN) could all affect the hot gas distribution, while stellar feedback is negligible \citep[e.g.][]{bigwood_case_2025}. However, the key mechanisms driving the emergence of X-ray bright and faint populations remain an open question. This paper addresses this issue by exploring several processes potentially responsible for these distinct populations, providing new insights into their formation and evolution.
\par
The AGN feedback plays a central role in shaping the galaxy luminosity function and regulating the thermal and dynamical properties of halos \citep[e.g.][]{planelles_role_2014}. Supermassive black holes (SMBHs) at the centres of galaxies can inflate large energetic bubbles that couple with the ICM and IGrM, pushing the inner gas to great distances. Additionally, outflows and jets inject vast amounts of energy into the IGrM, preventing gas reservoirs from cooling efficiently and suppressing star formation \citep{mcnamara_heating_2007, gaspari_linking_2020}. Shock fronts distribute this heat throughout the environment, and in galaxy groups, the total feedback energy may become comparable to or even exceed the halo's binding energy \citep[see an extensive review][]{eckert_feedback_2021}. While in galaxy clusters, this effect invests only the core region, in low-mass halos it extends to the virial scale \citep[e.g.][]{sorini_how_2022, angelinelli_mapping_2022, angelinelli_redshift_2023, ayromlou_feedback_2023}. State-of-the-art galaxy evolution models universally incorporate AGN feedback. Yet different implementations can yield drastically different predictions \citep{habouzit_supermassive_2021, schaye_flamingo_2023}, varying from exceedingly hot gas-rich groups to systems devoid of gas \citep[with a direct effect on the X-ray appearance of these systems on a global scale, see][]{popesso_x-ray_2024}. 
\par
In parallel, galaxy groups are structures embedded in a large-scale cosmic web, where gas is not only confined within the virialised halos but also extends into surrounding filaments in different phases \citep{eckert_warmhot_2015, graaff_probing_2019,tanimura_first_2020, galarraga-espinosa_properties_2021,  gouin_gas_2022, gouin_soft_2023, zhang_srgerosita_2024, dietl_discovery_2024}. These filaments act as gas reservoirs, locking up material for future accretion onto the central halo \citep{romano-diaz_zomg_2017}. The interaction between the group and these cosmic structures influences its hot gas content, X-ray properties, and overall dynamical state \citep[e.g.][]{manolopoulou_environmental_2021, damsted_axes-sdss_2024, cui_hyenas_2024}. For example, recent studies have hypothesised that ``fossil groups'' -- large X-ray halos associated with a few galaxies -- have their origin due to their position far from the nodes and dense environments of the cosmic web \citep{zarattini_fossil_2022, zarattini_where_2024}. The relative isolation they live in causes their galaxy members to merge undisturbed due to dynamical friction, resulting in the formation of massive halos near the base of the gravitational potential \citep[e.g.][]{donghia_formation_2005, jones_nature_2003} with very symmetrical appearances and high mass concentrations \citep{khosroshahi_scaling_2007,buote_unusually_2017, buote_extremely_2019, runge_unusually_2022}. This isolated evolution also explains the magnitude gaps between the first and N-th brightest galaxies (N-th can be second, and fourth depending on the studies) that characterise them \citep{aguerri_properties_2021}.
\par
In this work, we leverage the Magneticum Pathfinder simulations \citep{hirschmann_cosmological_2014, dolag_magneticum_2025}, a suite of state-of-the-art cosmological hydrodynamical simulations, to investigate the physical mechanisms behind the X-ray selection of galaxy groups. Magneticum provides a powerful framework for exploring how different assembly histories (i.e. halo accretion history) and feedback mechanisms lead to variations in gas content, BH growth, and X-ray luminosity. Unlike other cosmological simulations \citep[e.g. SIMBA, Eagle, and IllustrisTNG; see][]{habouzit_supermassive_2021}, Magneticum reproduces the AGN luminosity function well down to $z=2$ \citep{hirschmann_cosmological_2014, steinborn_cosmological_2018, biffi_agn_2018}, which is responsible for most of the X-ray cosmic background, and recovers several scaling relations obtained with eROSITA at the group and galaxy mass scales \citep{bahar_erosita_2022, zhang_hot_2024}. Magneticum achieves this without an explicit calibration on observed scaling relations down to the group-scale halos, in contrast to other state-of-the-art simulations, such as the IllustrisTNG project \citep{pillepich_simulating_2018, nelson_first_2018}, SIMBA \citep{dave_simba_2019}, and FLAMINGO \citep{schaye_flamingo_2023}. Thus, we argue that the Magneticum suite is an excellent tool to infer predictions of the hot gas in galaxy groups and investigate the detectability of such objects. By analysing synthetic X-ray observations and tracking the evolution of key properties such as gas fraction, environment, and SMBH activity, we aim to uncover the driving forces behind the X-ray selection of groups. Our findings may have important implications for the interpretation of X-ray surveys, the study of galaxy evolution, and the development of future observational strategies.  
\begin{figure*}
    \centering    
    \subfloat{\includegraphics[width=0.43\textwidth]{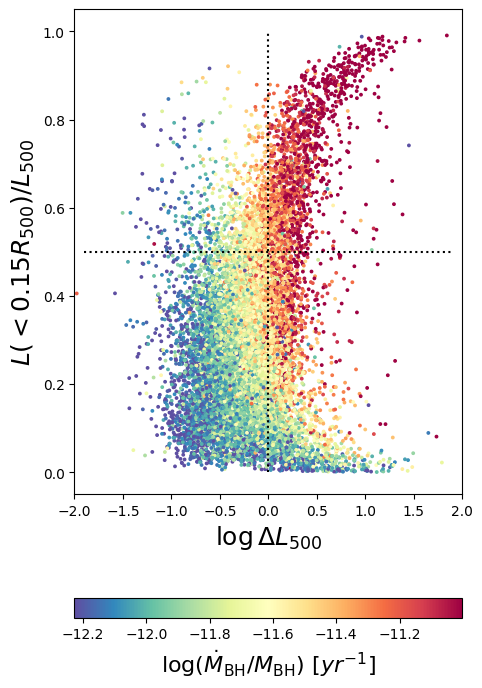}} 
    \subfloat{\includegraphics[width=0.43\textwidth]{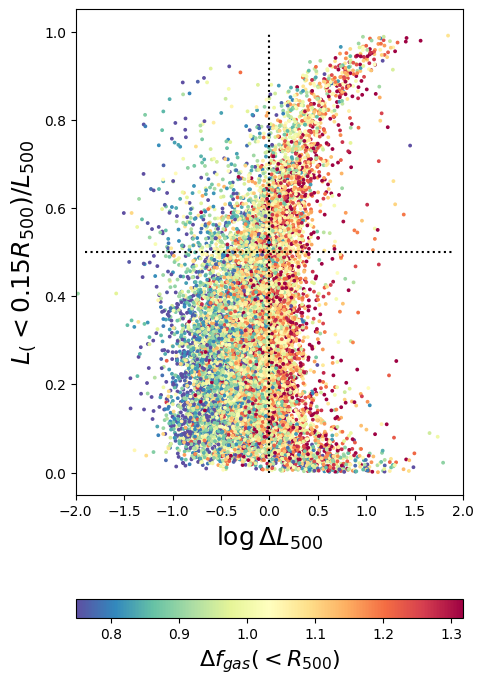}}  
    \caption{Ratio of the X-ray luminosity (intrinsic $0.5-2.0$ keV) of IGrM only within $0.15 R_{500}$ and $R_{500}$ as a function of the residual \deltaL. The latter is the residual distance between $L_{500}$ and the \cite{lovisari_scaling_2015} scaling relation. Left panel: We colour-coded the points based on the rescaled accretion rate of the central BH in the host halo. Right panel: We colour-coded the points according to the rescaled hot gas fraction in the halo. Both colour bars encompass the 5--95th percentile. See the main text for details. }
    \label{fig:centralL}
\end{figure*}

\begin{figure*}
    \centering    
    \subfloat{\includegraphics[width=0.8\textwidth]{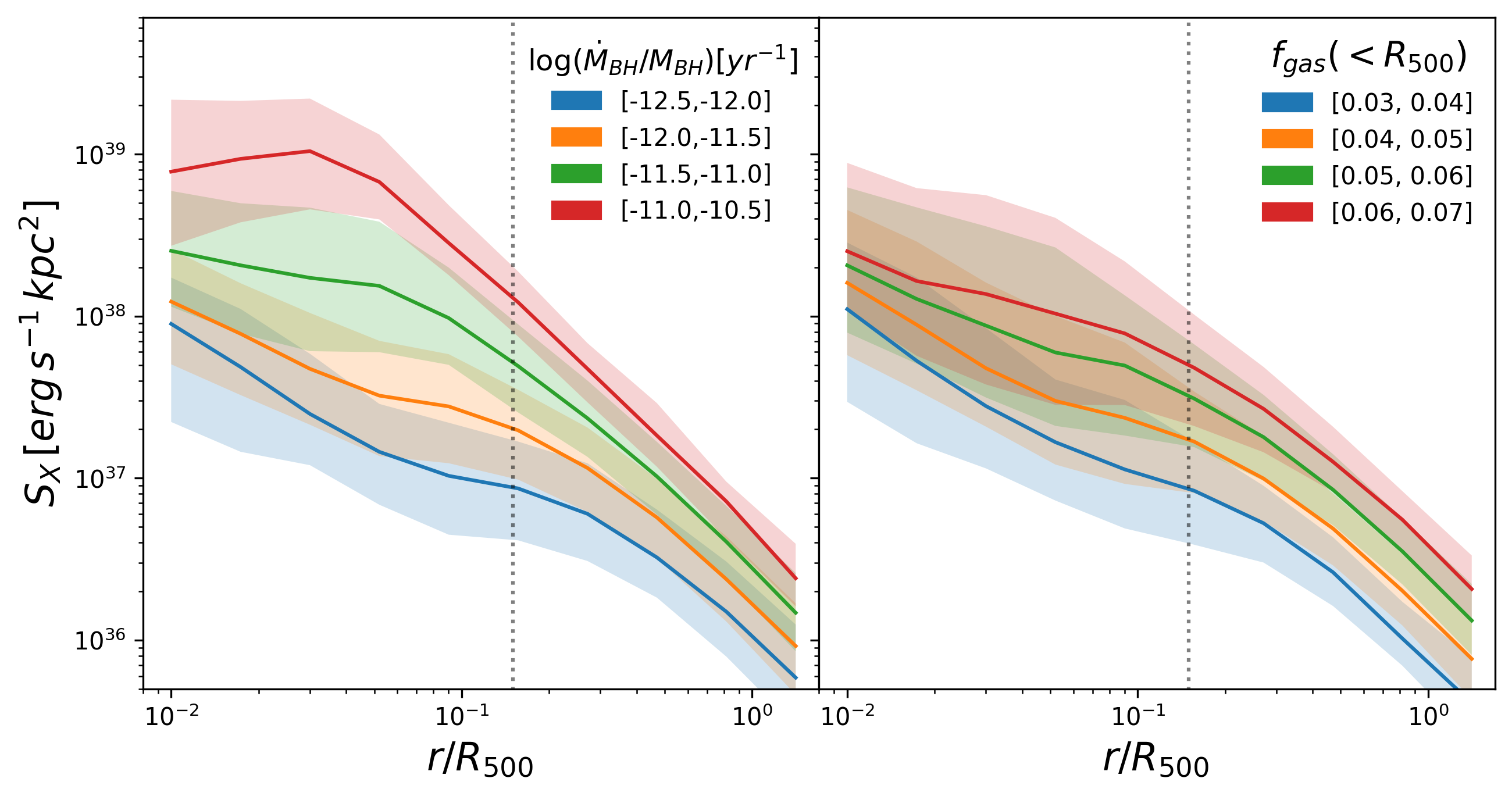}} 
    \caption{Median surface brightness profiles and 16--84th percentile of the X-ray emission (rest-frame $0.5-2.0$ keV) of the IGrM in groups in the halo mass bin $M_{500} = 10^{13}-10^{13.5} M_{\odot}$. The sample is divided into bins of the rescaled accretion rate of the central BH in the host halo (left panel) and hot gas fraction (right panel). We plot a dotted line corresponding to $0.15 R_{500}$ to separate the central from the external contribution given in Fig.~\ref{fig:centralL}.}
    \label{fig:centralS}
\end{figure*}

\section{Simulations}
This analysis is based on synthetic data from the X-ray lightcone described in \cite{marini_detecting_2024, Marini2025_erratum, marini_detecting_2025} and the original \textit{Box2/hr} at $z=0$ of the Magneticum set. The Magneticum Pathfinder simulation\footnote{\url{http://www.magneticum.org/index.html}} is an extended set of the state-of-the-art cosmological hydrodynamical simulations carried out with P-GADGET3 \citep[an updated version of the public GADGET-2 code][]{springel_cosmological_2005}. Key advancements include a higher-order kernel function, time-dependent artificial viscosity and artificial conduction schemes \citep{dolag_turbulent_2005,beck_improved_2016}.
\par
For this study, we use \textit{Box2/hr} which follows the evolution of a $352^3 h {^{-3}}$ cMpc$^{3}$ cosmological volume with $2\times1584^{3}$ particles at the mass following resolutions: $m_\mathrm{DM} = 6.9\times10^8 h{^{-1}}$~M$_{\odot}$ and $m_\mathrm{gas} = 1.4\times10^8 h^{-1}$~M$_{\odot}$. The Plummer equivalent length for the DM particles corresponds to $\epsilon=3.75 h^{-1}$ kpc, whereas gas, stars, and BH particles retain $\epsilon=3.75 h{^{-1}}$~kpc, $2 h {^{-1}}$~kpc and $2 h^{-1}$~kpc at $z=0$, respectively.
The simulation adopts the WMAP7 cosmology \citep{komatsu_hunting_2010}: $\Omega_\mathrm{M}=0.272$, $\Omega_\mathrm{bar}=0.046$, $n_s=0.963$, $\sigma_8 = 0.809$ and $H_0=100 \, h$ km~s${^{-1}}$~Mpc${^{-1}}$ with $h=0.704$. All logarithms are in base 10. 
\subsection{Subgrid physics}
\label{subsec: 2.1}
Several subgrid models describe the unresolved baryonic physics of the simulations including radiative cooling \citep{wiersma_effect_2009}, a uniform time-dependent UV background \citep{haardt_modelling_2001}, star formation and stellar feedback \citep[i.e. galactic winds][]{springel_cosmological_2003}, metal (H, He, C, N, O, Ne, Mg, Si, S, Ca, Fe) and chemical enrichment due to stellar evolution \citep{tornatore_chemical_2007}. Models for SMBH accretion and AGN feedback are included following the prescription in \cite{springel_cosmological_2005,di_matteo_energy_2005, fabjan_simulating_2010, hirschmann_cosmological_2014}. Briefly, the total energy released by a BH is modelled according to its gas accretion rate $\dot{M}_{BH}$ via a fractional radiative efficiency $\epsilon_r$ and the speed of light $c$, as in the following
\begin{equation}
\label{eq:EnergyBH}
    \dot{E}_\mathrm{BH} =\epsilon_r \dot{M}_\mathrm{BH} \, c^2 .
\end{equation}
The mass accretion $\dot{M}_\mathrm{BH}$ assumes the Bondi-Hoyle-Lyttelton approximation \citep[][and references therein]{hoyle_effect_1939} 
\begin{equation}
    \dot{M}_\mathrm{BH} = \frac{4 \pi G^2 M_{BH}^2 \alpha \rho}{(c_s^2 + v^2)^{3/2}},
\end{equation}
where $\rho$ and $c_s$ are the density and sound speed of the gas, respectively, $v$ is the BH velocity, and $\alpha$ is a boost factor (typically set to 100) which accounts for the limitations to the sub-kpc resolution not resolved in cosmological simulations. The mass accretion is capped at the Eddington limit \citep{eddington_radiative_1916} to balance inward gravitational attraction and outward radiation pressure, thus $\dot{M}_\mathrm{BH} = \mathrm{min}(\dot{M}_\mathrm{BH}, \dot{M}_\mathrm{Edd})$. Typically, only a fraction $\epsilon_f$ of the total energy in Eq.~\ref{eq:EnergyBH} is coupled with the surrounding medium and distributed kernel-weighted to the surrounding gas particles in an SPH-weighted fashion. A second mechanism contributing to BH growth is BH--BH mergers, which trigger an AGN feedback event proportional to the mass increase. We discuss the impact of different subgrid models in Sect.~\ref{subsec:SMBH}

\subsection{The LC30 lightcone and X-ray emission}
Complete geometric and physical description of the lightcone is provided in \cite{marini_detecting_2024, Marini2025_erratum, marini_detecting_2025}, here we briefly summarise the main points. Figure~\ref{fig:LC30_xray} illustrates the distribution of groups and clusters (with halo mass\footnote{We define $M_{\Delta}$ as the mass encompassed by a mean overdensity equal to $\Delta$ times the critical density of the universe $\rho_c(z)$.} $M_{200}>10^{12.5} M_{\odot}$) colour-coded for their X-ray luminosity within $R_{500}$. The lightcone spans $30 \times30$ square degrees (LC30) and consists of five sub-cubes extracted from the most recent five snapshots in \textit{Box2/hr}. The projected distribution of photons ($0.5-2.0$ keV) of each sub-cube is illustrated in the subpanels of Fig.~\ref{fig:LC30_xray}. Halos and substructures in the simulations are identified in post-processing using a two-step approach: first, a Friend-of-Friend (FOF) algorithm is applied, followed by SubFind \citep{springel_populating_2001, dolag_substructures_2009} to catalogue the subhalos. The X-ray emission is modelled using PHOX \citep{biffi_observing_2012, biffi_investigating_2013, biffi_agn_2018, vladutescu-zopp_decomposition_2023} which simulates X-ray spectral emission for the gas particles, the BH, and the star particles (i.e. X-ray binaries) using spectral models from the XSPEC library \cite[v12;][]{arnaud_xspec_1996}. A fiducial collecting area $A_{fid}$ and exposure time $\tau_{fid}$ are initially assumed to generate a statistically significant number of photons from the spectra using a Monte Carlo approach. These photons are then blueshifted or redshifted based on their location and projection within the lightcone. For this analysis, we do not use the eROSITA mock observation \citep[similarly to][]{marini_detecting_2024, Marini2025_erratum}, instead we calculate the X-ray luminosities from the theoretical PHOX's photon lists. These are estimated in the rest-frame $0.5-2.0$ keV energy range. 
\begin{figure}
    \centering
    \includegraphics[width=\linewidth]{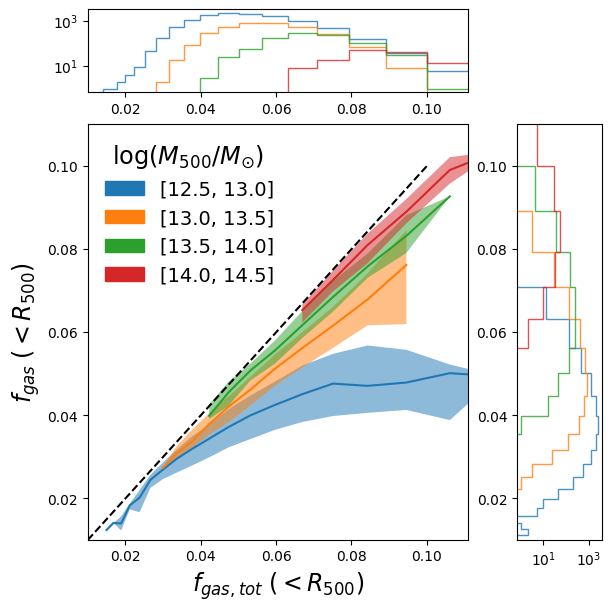}
     \caption{Hot gas fraction as a function of the total gas fraction within $R_{500}$. In the central panel, each solid line represents a halo mass bin $M_{500}$ with the associated 16--84th percentile. The dashed black line represents the 1:1 relation. In the lateral panels, we report the number counts of the halos of $f_\mathrm{gas,\, tot}$ (top) and $f_\mathrm{gas}$ (right) used to calculate the median and errors.}
    \label{fig:Fgas_Fhotgas}
\end{figure}
\section{Explaining the X-ray selection in galaxy groups}
Despite the paramount importance of galaxy groups from the astrophysical and cosmological point of view, the challenges in their detection have limited their studies. Many studies have been devoted to clarifying the X-ray selection function of X-ray surveys like eROSITA \citep{clerc_synthetic_2018,  liu_establishing_2022, seppi_detecting_2022, clerc_srgerosita_2024, marini_detecting_2024, Marini2025_erratum}. Here, we follow up on the work in \cite{marini_detecting_2024, Marini2025_erratum} which showed the selection function in X-ray surveys at the nominal depth of eRASS:4, highlighting a primary bias against galaxy groups with lower surface brightness and high core entropy. The authors also examined the scatter in the $L_{500}-M_{500}$ relation within Magneticum, calculating the residuals \deltaL{} from the \cite{lovisari_scaling_2015} $L_{500}-M_{500}$ scaling relation -- namely \deltaL{}$\equiv \log (L_{500}/L_{500,\, BF})$, where $L_{500}$ is the intrinsic luminosity and $L_{500, \, BF}$ is the best-fit value -- given that it provides the smallest residuals among the literature scaling relations compared in LC30 \citep{marini_detecting_2024, Marini2025_erratum}. The simulation indicates that the groups in the upper envelope of the relation exhibit a more dense gas distribution in the core (i.e. $0.15R_{500}$) than those in the lower envelope within a fixed halo mass bin.
\par
This section investigates whether the central hot gas\footnote{Here, and throughout the rest of the paper, we refer to hot gas as all the non-star-forming gas particles in the simulations with temperatures above $5\times 10^5 K$. Additionally, we use the subscript ``$f_\mathrm{gas}$'' to refer to hot gas and ``$f_\mathrm{gas, \, tot}$'' for both hot and cold gas.} concentration is the sole factor potentially enhancing or diminishing the total X-ray emission in galaxy groups and clusters in Magneticum. Under this hypothesis, the central region (e.g. within $0.15 R_{500}$) should account for most of the total X-ray emission and should define cool core systems \citep{gonzalez_villalba_how_2025}. We test this in Fig~\ref{fig:centralL}, where we illustrate the ratio between the central and total X-ray luminosity from the IGrM only (no XRB or AGN) of individual groups and clusters within LC30 at fixed halo mass as a function of \deltaL. To account for the halo mass dependence in the $L_{500}-M_{500}$ relation, we maintain the \deltaL{} definition which measures the distance between the halo's measured X-ray luminosity $L_{500}$ (in rest-frame $0.5-2.0$ keV) from the best-fit power-law scaling relation ($L_{500, \, BF}$) given in \cite{lovisari_scaling_2015} scaling relation \citep[see][]{marini_detecting_2024, Marini2025_erratum}. Values of \deltaL{} larger than zero correspond to over-luminous galaxy groups and vice-versa. Additionally, we define a luminosity concentration as the ratio between the luminosity calculated within $0.15 R_{500}$ and $R_{500}$. In Fig.~\ref{fig:centralL}, we quantify the importance of the central BH activity (left panel) and the hot gas content in $R_{500}$ (right panel) in shaping this distribution for objects with $M_{500}>10^{12.5}M_{\odot}$. The colour bar encompasses the 5--95th percentiles. To guide the eye, we overplot two dotted black lines over the \deltaL$=0$ and $L(<0.15 R_{500})/L_{500}=0.5$ relations.
\par
A large fraction ($\sim 45\%$) of the over-luminous halos (\deltaL{} $>0$) are centrally over-luminous, appearing in the top-right-hand region of the plot. This central emission is directly paired with BH activity, as the latter increases with high central gas content. A high gas fraction enables efficient BH feeding, leading to the release of thermal energy through AGN feedback. This energy returns to the otherwise cold star-forming gas, heating it, slowing down (if not quenching) star formation, and increasing the total hot gas mass within the halo. This process further boosts the X-ray emissivity, strengthening the relation with \deltaL. As illustrated by \cite{vladutescu-zopp_radial_2024}, this effect is not limited to extreme AGN cases and can hold for faint ones. Rather, it reflects the broader connection between central gas concentration, the central BH, and the host halo properties. However, this dependence is negligible for less centrally bright halos -- those with $L(<0.15R_{500})/L_{500} < 0.5$ -- which make up $\sim 55\%$ of the over-luminous population, since the AGN influence does not significantly extend to large radii. The distribution and colour-coding of these points depend on the subgrid model and parameters included in the simulation describing the BH accretion efficiency and thermal interaction with the surrounding plasma.
\par
Next, we examine the dependence on the hot gas fraction $f_\mathrm{gas}$ within $R_{500}$. We define $f_\mathrm{gas}(<R_{500})=M_{\mathrm{gas}, \, 500}/M_{500}$, the ratio of the hot gas in $R_{500}$ to the total halo mass in the same radius. Since this fraction varies with halo mass \citep[e.g.][]{angelinelli_mapping_2022}, we define the residuals hot gas fraction $\Delta f_\mathrm{gas} (<R_{500})$ as the deviation from the median hot gas fraction $\langle f_\mathrm{gas}(<R_{500})\rangle$ for a given group or cluster mass. Therefore, $\Delta f_\mathrm{gas}(<R_{500}) = 1$ corresponds to groups with median gas fractions for their host halo mass, whereas $\Delta f_\mathrm{gas}(<R_{500}) > 1$ (or $<1$) reflects halos with higher (lower) hot gas fractions than expected.
\par
We find that the hot gas fraction remains relevant across the full range of luminosity concentration $L(<0.15 R_{500})/L_{500}$, implying that some bright sources correspond to halos with high hot gas fractions but less peaked central gas concentrations (see the lower right panel of Fig.~\ref{fig:centralL}). This variation in luminosity distribution reflects the diversity of thermodynamical states in galaxy clusters and groups, which can arise from differing formation histories, merger activity, and AGN feedback efficiency. Halos with higher hot gas fractions but less peaked central concentrations likely represent systems where AGN feedback or merger history have redistributed gas, suppressing the formation of a strongly peaked core. Conversely, halos with both high $f_\mathrm{gas}$ and peaked central concentrations may indicate more relaxed systems with efficient gas cooling that fuels both star formation and BH activity.
\par
This argument is further supported by Fig.~\ref{fig:centralS}. We focus on a single halo mass bin, $M_{500}=10^{13}-10^{13.5} M_{\odot}$, and present the median surface brightness profiles ($0.5-2.0$ keV) with 16–84th percentiles, binned by rescaled BH accretion rate (left panel) and hot gas fraction (right panel). The vertical dotted black line marks $0.15 R_{500}$, distinguishing the central from the outer IGrM contribution, as shown in Fig.~\ref{fig:centralL}.
\par
The central region exhibits a distinct trend for systems with active BH accretion at fixed mass, while variations in the hot gas fraction have a more pronounced effect in the outer regions. Notably, these effects are not mutually exclusive—high central and outer gas fractions can coexist, or only one may dominate, depending on the timescales required for gas mixing.
\par
In conclusion, the hot gas distribution in galaxy groups plays a key role in their X-ray detectability. X-ray bright systems retain a higher hot gas fraction than faint ones at the same total mass, but different gas distributions influence detectability in distinct ways.

\begin{itemize} \item Groups with $L (<0.15 R_{500})/L_{500}>0.5$ exhibit high central brightness, retaining most of their hot gas at the centre. This triggers AGN feedback, which heats the gas, preventing cooling and increasing the total hot gas mass. Consequently, the overall gas density and X-ray emissivity rise.
\item Groups with $L (<0.15 R_{500})/L_{500}<0.5$ have high surface brightness without a strong central peak. These systems generally maintain higher hot gas fractions than others of the same total mass.
\end{itemize}

Overall, X-ray bright halos -- regardless of gas distribution -- retain higher gas fractions at least up to $R_{500}$, while X-ray faint systems are baryon-depleted. In the following sections, we investigate the physical mechanisms driving this dichotomy.
\begin{figure*}
    \centering    
    \subfloat{\includegraphics[width=0.45\textwidth]{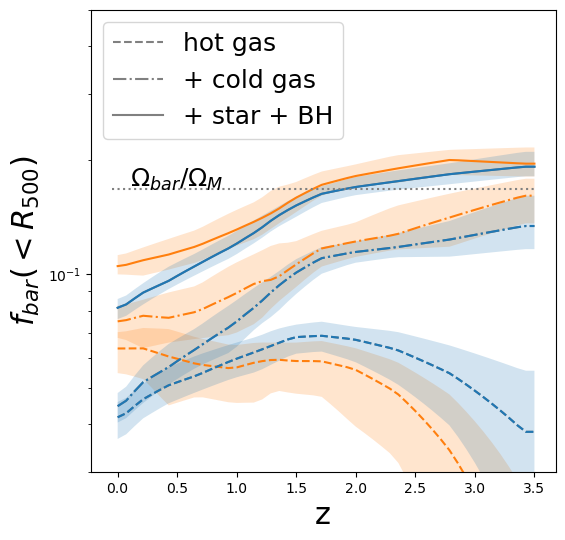}}     \subfloat{\includegraphics[width=0.45\textwidth]{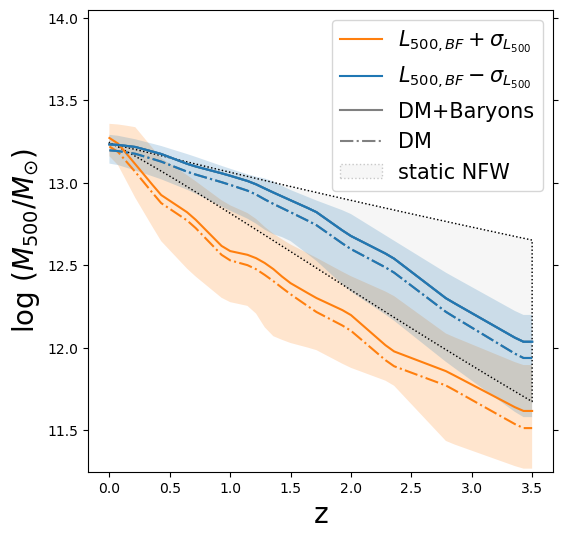}} 
    \caption{Evolution of several halo properties for a sample of bright and faint halos in the halo mass range $M_{500} = 10^{13}-10^{13.5} M_{\odot}$ at $z=0$. Left panel: Median baryonic fraction (solid lines) across different redshifts. We also report the hot gas contribution (dashed) and total gas (dashed-dotted). The fractions are calculated within $R_{500}$ of the group at the corresponding redshift. Right panel: Median halo mass as a function of the simulation redshift. The dashed-dotted lines represent the contribution only from the DM. We also report the pseudo-evolution of a static NFW profile in the range of concentration of our simulated halos $3<c_{500}<10$. The legend is common in all panels: the halo samples are selected from the original Magneticum cosmological box (i.e. \emph{Box2/hr}). The shaded regions mark the 16--84th percentile region.  }
    \label{fig:growth_z_13-13.4}
\end{figure*}

\begin{figure}
    \centering
    \subfloat{\includegraphics[width=0.45\textwidth]{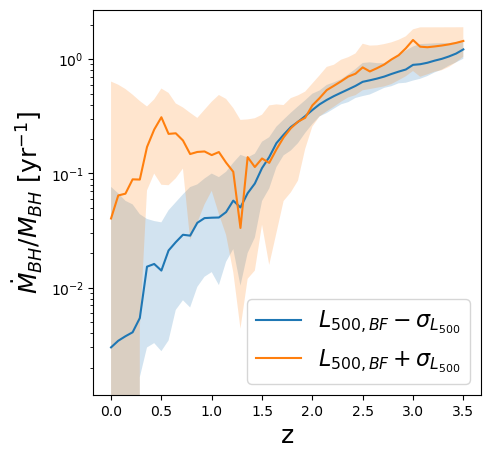}} 
    \caption{Evolution of thermal energy released by the central AGNs in the bright and faint samples from Fig.~\ref{fig:growth_z_13-13.4} in the halo mass range $M_{500} = 10^{13} - 10^{13.5} M_{\odot}$. The feedback energy is estimated from the BH accretion rate and its mass. The shaded bands report the 16--84th percentile of the distribution.}
    \label{fig:MdotBH_MBH_z_13-13.4}
\end{figure}
\subsection{Build-up of the gas content}
\label{sec:4}
A centrally peaked hot gas distribution can only partially explain the selection effect in X-ray emission at group scales. More generally, the total hot gas content within the virial radius determines its detectability. The build-up of this component is influenced by both gravitational and non-gravitational processes (e.g. mergers, AGN feedback, and star formation), as well as phase mixing, all of which can significantly contribute to the scatter in the $M_{500}-f_\mathrm{gas}$ relation and, in turn, the $L_{500}-M_{500}$ relation.
\par One might wonder how representative the hot gas fraction is of the total gas content in a group or cluster of galaxies. To address this, we analyse the expected hot gas fraction (hot gas mass over halo mass within $R_{500}$) as a function of the total gas fraction (hot and cold gas over halo mass within $R_{500}$) in Fig.~\ref{fig:Fgas_Fhotgas}. It is worth mentioning that Magneticum's predictions on the $f_\mathrm{gas}-M_{halo}$ relation are in excellent agreement with the median and scatter of the hot gas fraction from eROSITA stacked groups \citep{popesso_hot_2024} and the XXL groups and clusters with halo mass measurements from weak lensing \citep{akino_hsc-xxl_2022}. We divide the LC30 sample into logarithmic halo mass bins of $M_{500}$, showing their scatter (coloured shaded regions), and plot the 1:1 scaling relation in dashed black. If the relation follows the 1:1 line, this implies that all gas in the halo is hot. In the lateral panels, we plot the number counts of halos in the bins we used to calculate the median and error.
\par We deduce that if a halo has a low total gas fraction compared to haloes with a similar mass, most (if not all) of it is in the hot phase. The gravitational potential well compresses the hot plasma, increasing its kinetic energy, which is later converted into thermal energy. As a result, larger halos tend to heat more gas. In the clusters regime, this process converts most of the gas into a hot, X-ray-bright component. However, in less massive halos, the colder (star-forming) gas component can survive longer within $R_{500}$. Consequently, a larger total gas fraction does not necessarily imply a well-mixed hot halo. We argue that measuring only the X-ray-emitting gas in the group regime ($M_{500} < 10^{13.5} M_{\odot}$) could lead to a significant underestimation of the total gas fraction (up to 50\%) if this effect is not considered -- at least in the $f_\mathrm{gas, \, tot}>0.05$ regime of our simulations. This result is robustly predicted by the number of halos that are estimated to have $f_\mathrm{gas, \, tot}$ in the regime $0.05-0.10$, which represent $\geq 50\%$ of the groups in $M_{500} < 10^{13.5} M_{\odot}$ (see top histograms). In this regard, recent studies on the gas fraction have called for stronger AGN feedback than most hydrodynamical simulations (but not Magneticum, see \citealt{popesso_average_2024}) based on evidence collected from the cosmic shear, X-ray, and SZ measurements \citep[e.g.][]{preston_non-linear_2023, amon_consistent_2023, bigwood_weak_2024, mccarthy_flamingo_2024, hadzhiyska_evidence_2024, ferreira_x-ray-cosmic-shear_2024, la_posta_xy_2024}. However, we highlight here the need to account for complementary techniques to include cold gas (e.g. measuring the MgII covering factor; \citealt{anand_characterizing_2021}) in Milky Way-mass halos. 
\par
To investigate which factors play the most significant role in the cosmic structure's evolution, we track a sample of halos and their most massive progenitor down to $z\simeq3.5$. Several of these halos are also part of LC30; however, relying solely on this subset would be insufficient for robust statistical conclusions. Therefore, we randomly select $\sim 200$ halos from the lowest snapshot (i.e. snapshot 140, $z=0.0326$) of the original Magneticum \emph{Box2/hr} dataset and categorise them based on their location in the $L_{500}-M_{500}$ scaling relation. We define a faint subsample ($L_{500,\, BF}-\sigma_{500}$) and a bright subsample ($L_{500,\, BF}+\sigma_{500}$), selecting halos that lie outside the 1--$\sigma_{500}$ distribution from the best-fit mean relation at $z=0.03$. Notice that this definition of bright and faint does not necessarily hold at higher redshifts, where a faint object might have yielded a higher gas fraction and thus higher $L_{500}$. Each sample contains approximately 100 galaxy groups with halo masses $M_{500}$ ranging between $10^{13}-10^{13.5} M_{\odot}$ at $z\simeq0.03$.
\par
Figure~\ref{fig:growth_z_13-13.4} shows our results for the halo mass bin $M_{500} = 10^{13}-10^{13.5} M_{\odot}$; however, we remark these results also hold in different halo mass bins within $M_{500} = 10^{12.5}-10^{14.5} M_{\odot}$. We analyse the total baryonic fraction (left panel) and the halo mass (right panel) within $R_{500}(z)$ up to the maximum redshift when we can trace the parent halo, in most cases $z=3.5$. The bright sample is marked with orange, whereas the faint sample is in blue. We report the median and 16--84th percentile with the shaded area.
\par
At the group scales, the baryonic fraction is expected to decrease from the cosmic value due to a dominating AGN feedback \citep[see, e.g.][]{sorini_how_2022, ayromlou_feedback_2023}. Indeed, at early times the distributions of the bright and faint samples are consistent with each other (and to the cosmic value $\sim 0.168$ reported with the dotted line). However, around $z\simeq 1-1.5$, we observe a consistent splitting in their evolution that leads to different total baryonic fractions at $z=0$. This epoch is key to the different evolutionary paths. The bright sample reaches a plateau in the total baryonic fraction, whereas the faint sample keeps decreasing. We point out the different timescales affecting all components: in all mass bins, the divergence happens firstly for the cold gas + star + BH component, and only later it affects the hot gas component. 
\par
Furthermore, the mass accretion history is strikingly different in the two samples (right panels in Fig.~\ref{fig:growth_z_13-13.4}). We overplot the self-similar evolution of a DM halo in the given mass range at $z=0$ expected in the range of concentration $c_{500}$\footnote{The DM concentration $c_{\Delta}$ defines how centrally concentrated the DM is in a halo. Analytically it reads $c_{\Delta} = R_{\Delta}/R_s$ where $R_s$ is the scale radius in an NFW profile \citep{navarro_universal_1997}.} of our simulated halos $3<c_{500}<10$ \citep{diemer_pseudo-evolution_2013, diemer_colossus_2018}. Faint halos typically gain most of their mass early on, with later phases marked by gas depletion due to star formation, AGN activity, and stellar feedback. Interestingly, the halo mass growth experiences a slowdown in the faint sample at the epoch when the baryon fraction begins to decline. Such a decline can indeed be connected with a general failure in replenishing the gas content in the halo, which will reduce in the future due to various feedback and star formation. However, their overall evolution is consistent with the pseudo-evolution of a stationary NFW density profile (shown in Fig.~\ref{fig:growth_z_13-13.4}), changing due to the evolution of a halo mass defined by a density threshold $\Delta = 500$ that decreases with time. In contrast, bright halos exhibit steady exponential accretion at all times, removing the $1$ dex difference in halo mass originally present at $z=3.5$. A visual inspection of the bright halos at this epoch reveals that they undergo major mergers, whereas the faint ones exhibit very few substructures and display a regular, symmetric X-ray emission around the central galaxy --reminiscent of fossil groups \citep{aguerri_properties_2021}.
\par
Another way of quantifying such difference is to estimate the median redshift of formation of the two samples, defined as the median redshift at which half of the halo mass is in place. These correspond to $z_{L_{500, \, BF}+\sigma_{500}}= 1.2$ and $z_{L_{500, \, BF}-\sigma_{500}}= 2.1$, respectively. 
\par
Therefore we can assume that the difference in the X-ray emission is consistent with the availability of the hot gas in the halo, which depends on the halo's replenishment opportunities. In the next two sections, we investigate the role that the central SMBHs and the environment play in replenishing or depleting these halos. 
\begin{figure*}
  \centering
  \subfloat{\includegraphics[width=0.9\linewidth,trim={0 1.65cm 0 0},clip]{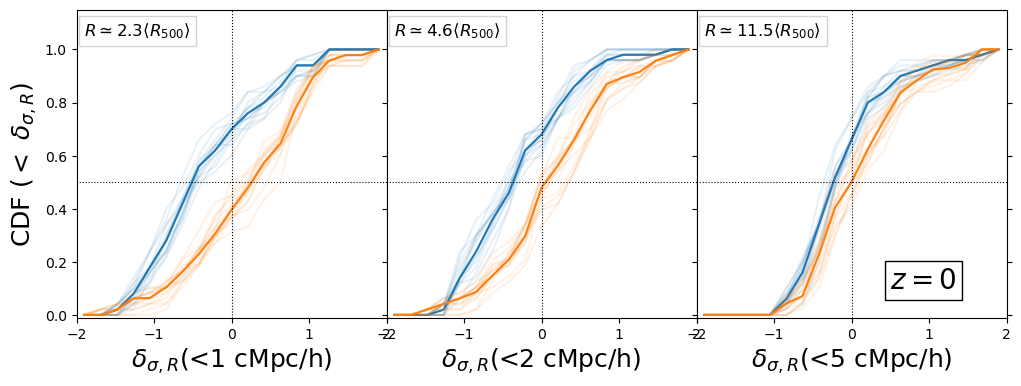} } \\
   \subfloat{\includegraphics[width=0.9\linewidth,trim={0 1.65cm 0 0.2cm},clip]{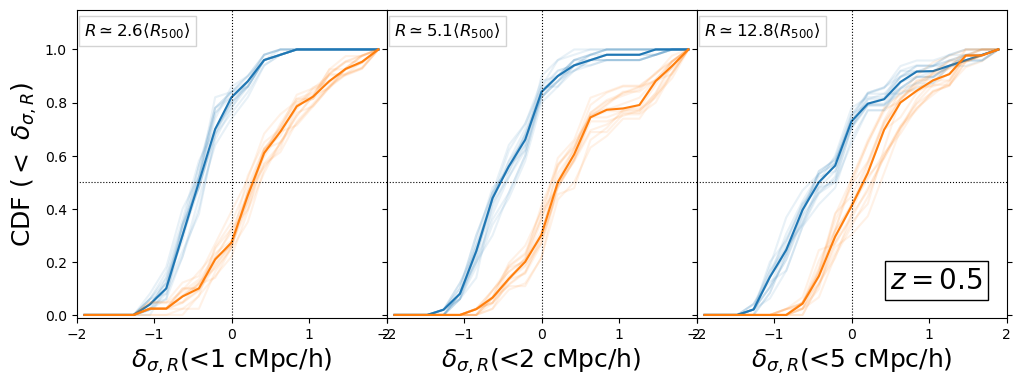}} \\
  \subfloat{\includegraphics[width=0.9\linewidth]{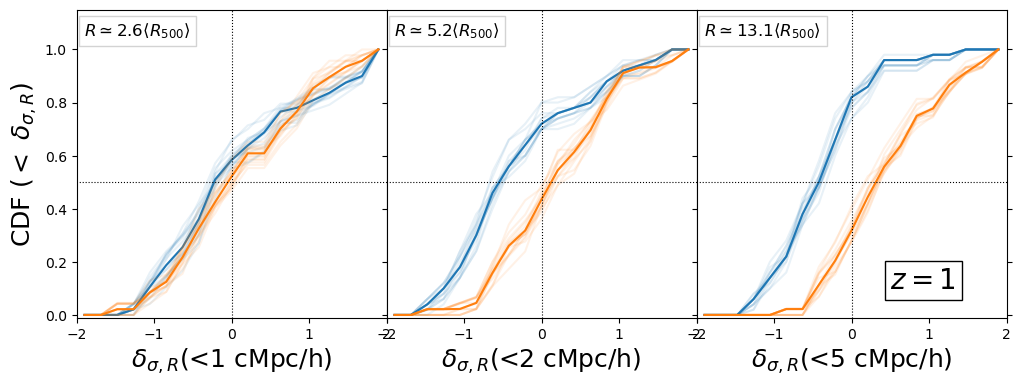}} \\
\caption{Cumulative density distribution of the normalised density contrast for bright (orange) and faint (blue) halos at three redshifts. The solid lines are the median of the bootstrapped sample (shaded lines) for halos within $10^{13}-10^{13.5} M_{\odot}$ selected at $z=0$. From left to right, the normalised density contrast is calculated for different radii, namely 1, 2, and 5 cMpc/h. In the top-left corner of each panel, we also report its correspondence to the median $R_{500}$ in the sample. We overplot a vertical and horizontal line to guide the eye. } \label{fig:environment_z0}
\end{figure*}
\subsection{The role of the central SMBH}
The activity of the central and satellite SMBHs can severely impact halos across their cosmic time. While AGN feedback heats and displaces the gas, altering the halo's thermodynamic state \citep{eckert_feedback_2021}, many simulations \citep{steinborn_origin_2016, steinborn_cosmological_2018, castro_impact_2021} have shown it is insufficient to eject gas entirely from the halo at late times. In reality, most of the strong AGN activity episodes happen at the early phases of the halos assembly \citep[$z=2-3$, see for example][]{castro_impact_2021,ragagnin_simulation_2022, ayromlou_feedback_2023} which can cause an efficient displacement of the gas well beyond the virial radius at the group scales. On the other hand, DM is less sensitive to the gas displacement (and completely insensitive to the thermal feedback), leading to a continuous accretion even at later times.  \cite{biffi_agn_2018} and \cite{marini_detecting_2024, Marini2025_erratum} showed that Magneticum correctly reproduces the AGN luminosity function at $z=0$ and in LC30, respectively. Therefore, we expect the simulated AGN population to be representative of the real one.
\par
In this section, we investigate whether one or multiple uncommon AGN episodes in the faint sample at around $z\simeq0-1.5$ might have led to gas depletion within their virial radius. We exclude the energy feedback from stellar processes in this analysis, because of its minor contribution compared to AGN feedback at the redshifts we are interested in.
\par
Regrettably, detailed information on the BH accretion properties at each simulation timestep is unavailable for our simulation run. However, we can retrieve a rough estimate of its global energy injection in each halo with arguments derived from the AGN feedback scheme implemented in Magneticum that follows the structure described in \cite{di_matteo_energy_2005} and \cite{springel_cosmological_2005} with minor modifications implemented in \cite{hirschmann_cosmological_2014} and \cite{steinborn_refined_2015}. Figure~\ref{fig:MdotBH_MBH_z_13-13.4} shows the AGN activity estimated for the central SMBHs in the two samples. We plot the SMBH accretion rate $\dot{M}_{BH}$ renormalised for its mass $M_{BH}$. We quantify the BH accretion rate $\dot{M}_{BH}$ as the mass difference between two consecutive snapshots rather than the BH mass accretion rate saved in the simulation snapshot, which only measures the instantaneous accretion rate. 
\par
The BH evolution demonstrates that the feedback activity of the faint sample steadily declines over time, with no recent episodes accounting for the decrease of the gas content around $z\simeq 0-1.5$. In contrast, the bright sample exhibits a rise in AGN activity at the epoch when the baryonic fraction changes ($z\simeq 1-1.5$). If a major merger occurs at this redshift, the instantaneous availability of gas in the halo fuels the central SMBH, triggering a thermal energy release in the bright sample. The BH accretion is more efficient and less turbulent during the accretion of cold gas particles, although both cold and hot gas phases contribute to the feeding process. Because cold gas is accreted more efficiently than hot gas, \emph{(i)} it is quickly depleted and \emph{(ii)} this selective removal of low-entropy gas boosts the overall temperature \citep{mccarthy_2011_gas}. In response, the SMBH releases thermal energy, which, while insufficient to fully deplete the gas reservoir of the halo, significantly increases the hot gas fraction.
\par
In conclusion, the X-ray detectability of galaxy groups and clusters is shaped by the interplay between gravitational and non-gravitational processes. Bright halos maintain a steady accretion history and higher hot gas fractions due to sustained AGN activity and feedback that heat the gas without ejecting it. In contrast, faint halos evolve having lost the capability of reaccreating baryons, resulting in lower hot gas fractions and reduced X-ray brightness. We exclude that the AGN activity at late times can explain the depletion of baryons in faint galaxy groups. Alternatively, the past AGN activity might have displaced the gas in these halos that cannot (or have not yet been able) to re-accrete it. This complex interplay of processes introduces significant scatter in scaling relations like $L_{500}-M_{500}$ and highlights the importance of accounting for feedback and cooling histories in understanding the thermodynamic evolution of galaxy systems. 

\subsection{The role of the environment}
The role of the central SMBH and its feedback is often invoked to explain gas depletion at the group scales \citep[e.g.][]{ayromlou_feedback_2023}; however, Fig.~\ref{fig:growth_z_13-13.4}--\ref{fig:MdotBH_MBH_z_13-13.4} highlight the importance of mass accretion history in replenishing the gas reservoir of bright halos, rather than AGN-driven gas depletion at late times. This suggests that the ability of a halo to re-accrete gas from its surroundings plays a crucial role in shaping its thermodynamic properties. In this section, we investigate whether these evolutionary differences arise from the surrounding cosmic environment and whether halo assembly bias--the dependence of halo properties on their large-scale environment at fixed mass--contributes to the observed dichotomy.

\par To assess this, we examine whether bright and faint halos preferentially reside in over- or under-dense environments. We define the density contrast $\delta$ as

\begin{equation} \delta(< R) = \frac{\rho(< R) - \langle\rho\rangle}{\langle\rho\rangle} \end{equation}

where $\rho(< R)$ is the density within a sphere of radius $R$ around a halo and $\langle\rho\rangle$ is the mean cosmic density at the given redshift. To generalise the use of the density contrast across different radii, we introduce the normalised density contrast $\delta_{\sigma, R}$, given by

\begin{equation} \delta_{\sigma, R} = \frac{\delta (<R) - \langle\delta (<R)\rangle}{\sigma_{\delta,\, R} (<R)}. \end{equation}

Here, $\langle{\delta} (<R)\rangle$ and $\sigma_{\delta,\, R} (<R)$ represent the mean and standard deviation of the cosmic density contrast distribution, respectively. This formulation quantifies how much a halo’s density contrast deviates from the sample mean: a positive $\delta_{\sigma, \,R}$ corresponds to an over-dense environment, whereas a negative value indicates an under-dense region.

\par Figure~\ref{fig:environment_z0} shows the cumulative distribution function of the normalised density contrast for the bright and faint samples. The solid lines represent the median of the bootstrapped sample, while the shaded regions indicate the uncertainties. A striking difference emerges between the two populations, visible across all panels (from left to right, where $\delta_{\sigma, R}$ is computed at different radii). The X-ray bright sample is preferentially located in median-dense/over-dense environments, naturally explaining its enhanced ability to accrete gas over cosmic time. Conversely, the X-ray faint sample is consistently in the under-dense region of the diagrams at all $z$. Additionally, we find a statistically significant preference for higher richness (i.e. the number of gravitationally bound subhalos) in the bright sample, a trend that persists at higher redshifts.
\par
The observed environmental dependence strongly suggests an underlying halo assembly bias, in which the early growth history of a halo and its surrounding large-scale structure influence its baryon fraction and X-ray brightness at fixed halo mass. Moreover, the relatively relaxed early assembly history of X-ray faint systems aligns with the formation models of the commonly known ``fossil groups'' \citep{kundert_are_2017, kanagusuku_fossil_2016}.

\par
\begin{figure}
    \centering
    \includegraphics[width=\linewidth]{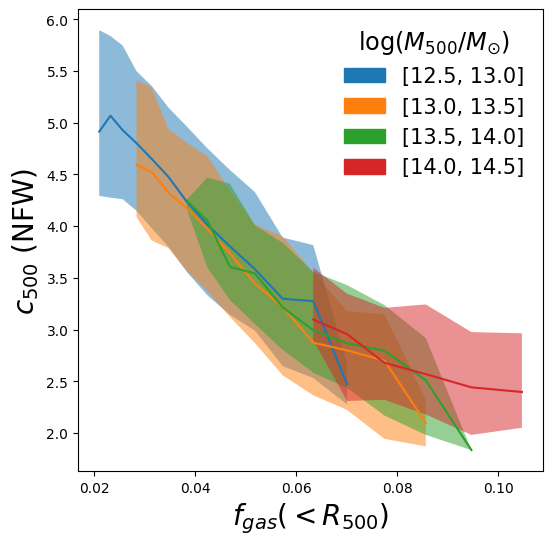}
    \caption{Median of the DM concentration parameter $c_{500}$ as a function of the hot gas fraction in $R_{500}$. Each line represents a different halo mass bin. The shaded band reports a 16--84th percentile. }
    \label{fig:c500_Fhotgas}
\end{figure}

 A natural consequence of sustained mass accretion in the bright sample and the relatively quiescent evolution of the faint sample is the impact on the structural properties of their DM halos, particularly their concentration. Seminal studies by \cite{wechsler_concentrations_2002} and \cite{gao_age_2005} have demonstrated a correlation between the concentration parameter $c_{\Delta} = R_{\Delta}/R_s$ (where $R_s$ is the scale radius in an NFW profile) and halo formation time. Halos that undergo an extended period of slow growth tend to have higher concentrations, while those experiencing recent mergers or large mass inflows develop lower concentrations \citep{wechsler_concentrations_2002, zhao_growth_2003}. To explore this further, we examine how the concentration parameter $c_{500}$ varies with hot gas fraction for halos from LC30 in Fig.~\ref{fig:c500_Fhotgas}. Since at fixed redshift and cosmology, the $c_{\Delta}-M_{\Delta}$ relation is in place, leading to numerous numerical calibrations \citep[e.g.][]{dutton_cold_2014}, we divide our sample into four mass bins and plot the running median and 16--84th percentile of $c_{500}$ as a function of the hot gas fraction within $R_{500}$. Across all mass bins, we observe a strong anti-correlation (a Spearman correlation of $\rho_{\{X, Y\}} = -0.61$): halos with higher gas fractions (which formed more recently) exhibit lower concentrations, consistent with a more recent and likely more chaotic mass accretion history. This trend is present across all mass ranges, further supporting the idea that environmental effects and assembly history shape the thermodynamic properties of halos at fixed mass.

\par 

\begin{figure*}
    
    \centering   
    \subfloat{\includegraphics[width=0.2965\textwidth]{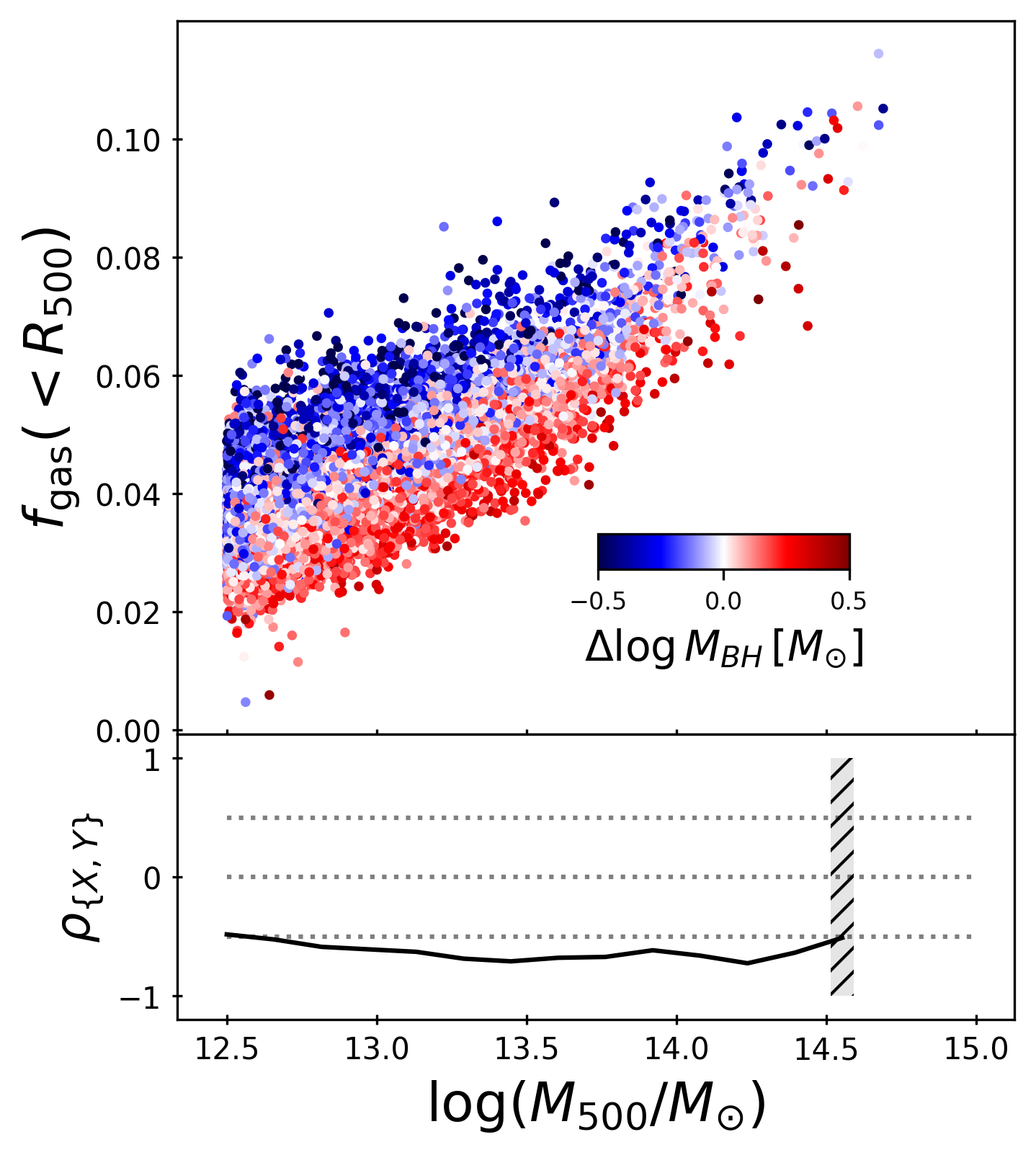}}  
    \subfloat{\includegraphics[trim={1.9cm 0 0 0}, clip, width=0.25\textwidth]{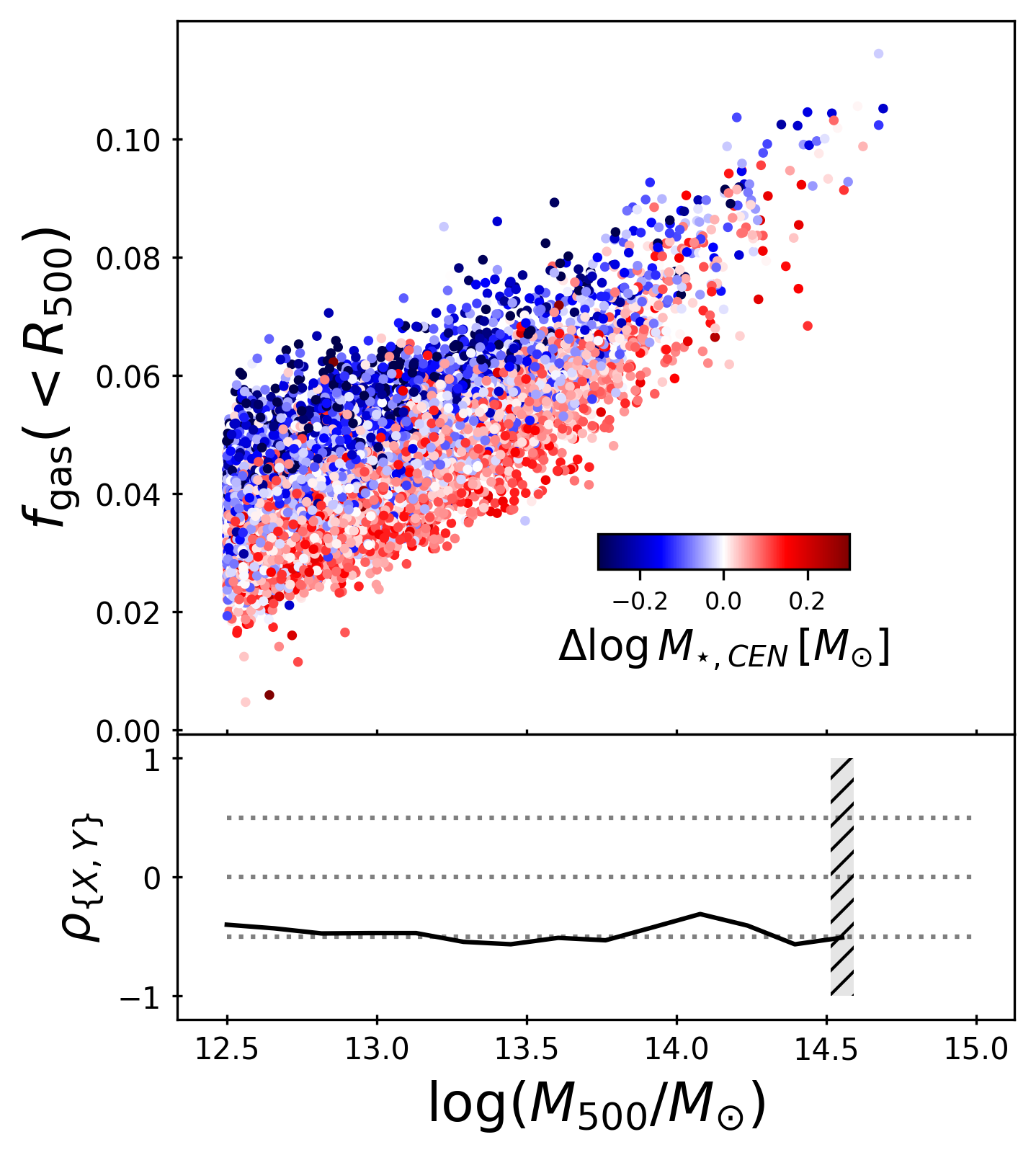}}  
    \subfloat{\includegraphics[trim={1.9cm 0 0 0}, clip,width=0.25\textwidth]{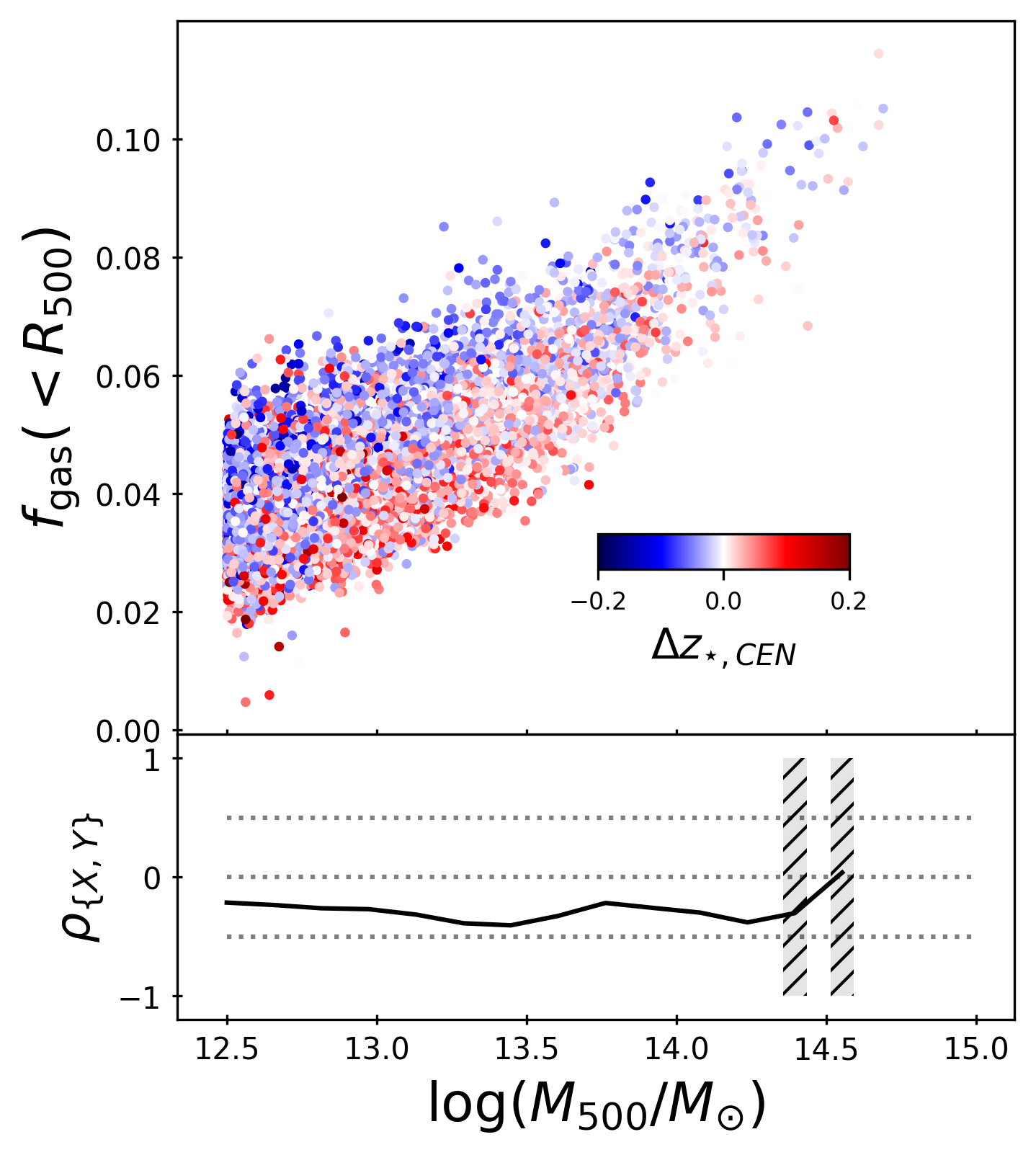}}\\ 
    \subfloat{\includegraphics[width=0.282\textwidth]{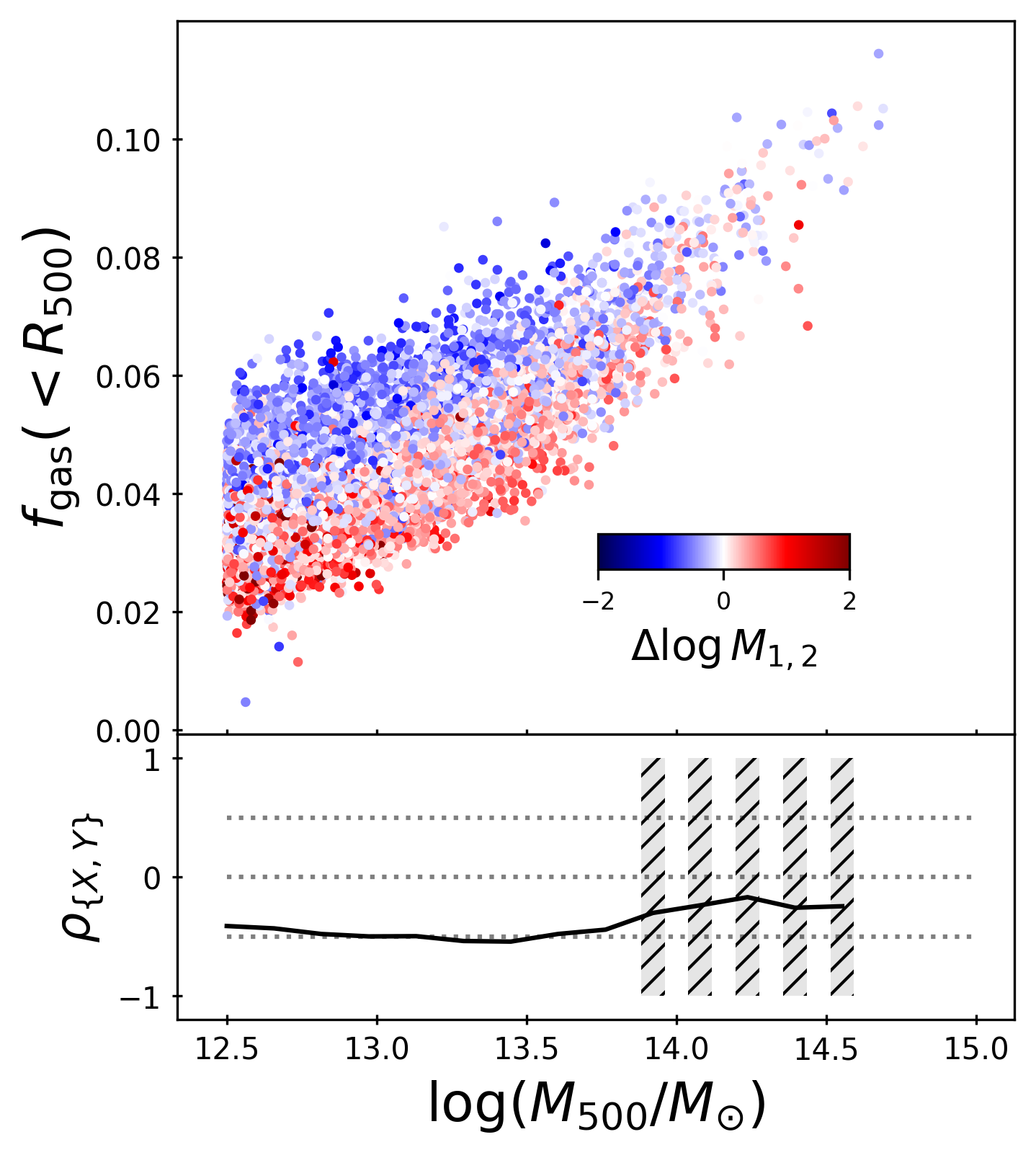}} 
    \subfloat{\includegraphics[trim={1.9cm 0 0 0}, clip,width=0.24\textwidth]{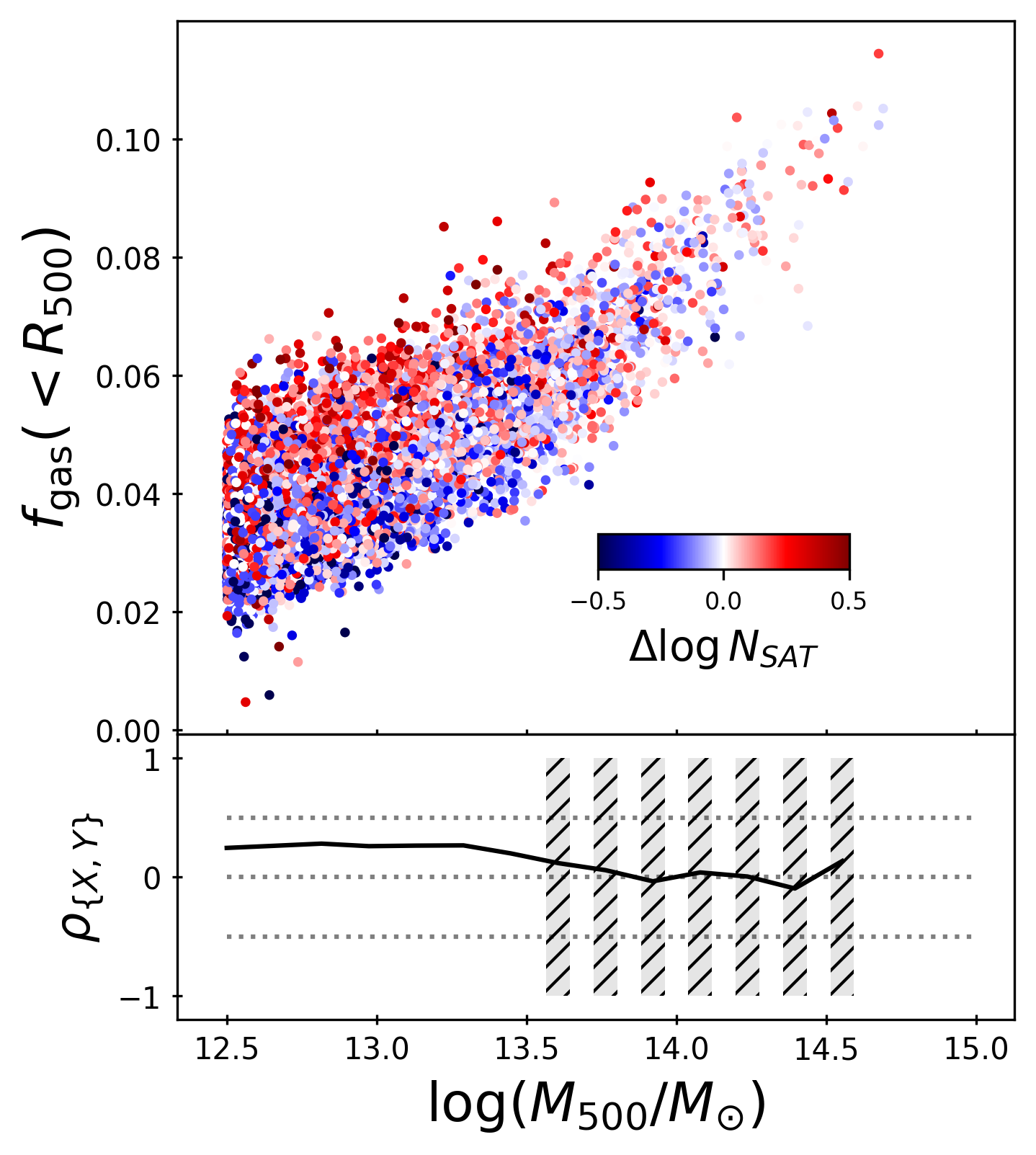}}
    \subfloat{\includegraphics[trim={1.9cm 0 0 0}, clip,width=0.24\textwidth]{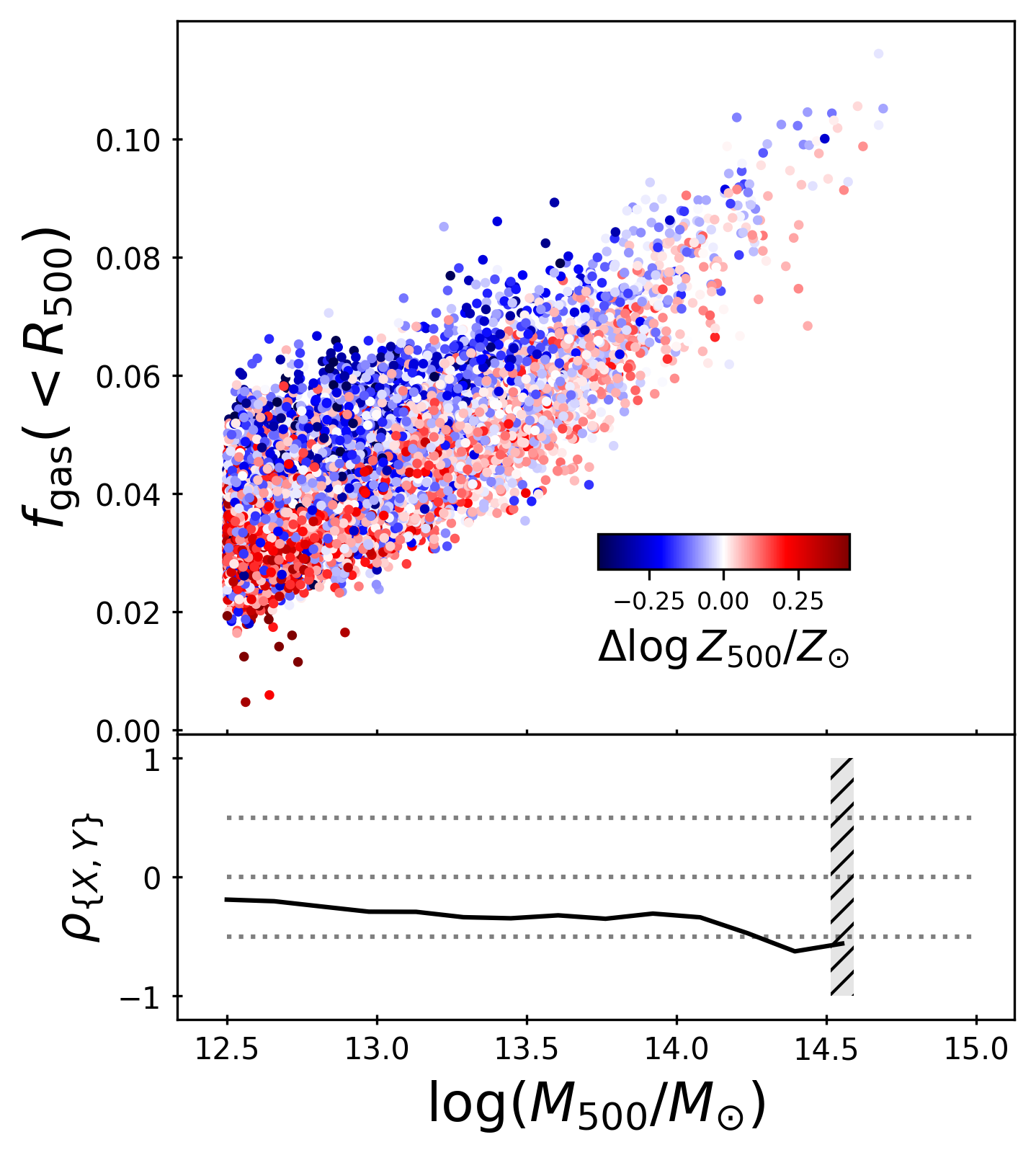}} 
    \subfloat{\includegraphics[trim={1.9cm 0 0 0}, clip,width=0.24\textwidth]{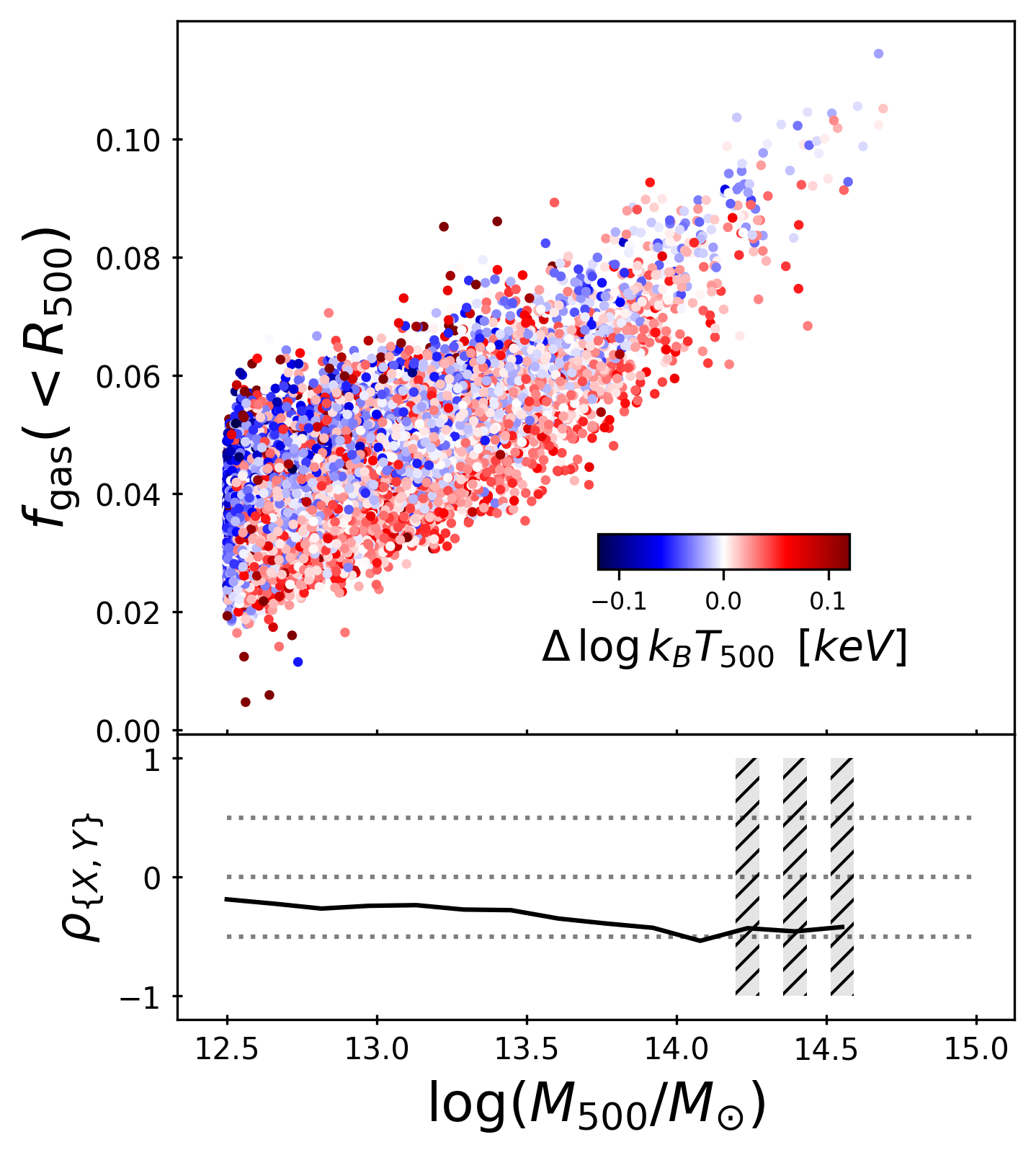}} 
    \caption{Hot gas fraction as a function of the halo mass. The points are colour-coded for the residuals about the median of different properties as a function of the halo mass. The properties are: the BH mass $\Delta \log M_\mathrm{BH}$, the BCG stellar mass $\Delta \log M_\mathrm{\star, CEN}$, the mean stellar formation redshift $\Delta z_\mathrm{\star, CEN}$, the magnitude gap $\Delta \log M_\mathrm{1,2}$, the richness $\Delta \log N_\mathrm{SAT}$, the total metallicity $\Delta \log (Z_{500}/Z_{\odot})$, and the total temperature $\Delta \log (k_B T_{500})$. In the bottom panels, we plot the Spearman coefficient between the hot gas fraction and the property in halo mass bins. The shaded band regions represent the bins where the p-value$<0.05$. The analogue plot with X-ray luminosity in the soft band is in Fig.~\ref{fig:appendix}. }
    \label{fig:Fhotgas_misc}
\end{figure*}

\subsection{Connection to global properties of groups and galaxies}
In this section, we analyse several global properties of groups and their central galaxies to highlight other possible observables affected by different evolutionary patterns. Given the tight connection between hot gas fraction and DM concentration (shown in Fig~\ref{fig:c500_Fhotgas}), we only show results as a function of the former, assuming that all the results inversely hold for the latter. We investigate:
\begin{itemize}
    \item the central SMBH properties, namely the BH mass ($M_{BH}$), its accretion rate and the number of past mergers;
    \item the fossilness ($\log M_{1,2}$), namely the difference in mass between the first and second most massive satellite in the simulation, considered as a proxy for the system's relative age in observations;
    \item the stellar mass ($M_{\star,\, CEN}$) and mean redshift of formation  ($z_{\star, CEN}$) for stars in the central galaxy or brightest cluster galaxy (BCG);
    \item the metallicity and temperature of the hot gas;
    \item the number of bound satellites.
    \end{itemize}
The choice of these properties comes from knowing how these parameters are expected to evolve if different assembly scenarios happen to halos at the same fixed halo mass at $z=0$. Furthermore, many of these properties are real observables that can be tested in observations. However, many (if not all) of these properties have a primary dependence on the halo mass which we need to take into account to support our hypothesis. We measure the residuals about the running median in halo mass bins for each of these properties and quantify the impact in the $f_\mathrm{gas}-M_\mathrm{halo}$ relation plotted in Fig.~\ref{fig:Fhotgas_misc}. As such, the colour coding in the top panels reflects this trend: positive values of each of the above properties correspond to an excess to the median value in the fixed halo mass bin, whereas negative values are below the median. In the bottom panels, we report the Spearman coefficient between the gas fraction and the property in bins of halo mass. The shaded grey areas represent regimes in which the p-value$<0.05$. We provide the same plots for the X-ray luminosity in the soft band (i.e. $0.5-2.0$ keV) in Fig.~\ref{fig:appendix}.

\subsubsection{The SMBH properties}
\label{subsec:SMBH}
The first panel in Fig.~\ref{fig:Fhotgas_misc} illustrates the trend of the $f_\mathrm{gas}-M_\mathrm{halo}$ relation when the central SMBH properties are considered. We colour-code the points with the residuals about the running median in halo mass bins of the BH mass $\Delta \log M_\mathrm{BH}$. We do not show the instantaneous BH accretion rate $\Delta \log \dot{M}_\mathrm{BH}$ or the number of BH--BH mergers $\Delta \log N_\mathrm{(BH-BH) merger}$, as they do not correlate. The strongest anti-correlations with the hot gas fraction are found for the BH mass ($\rho_{\{X, Y\}}\leq -0.5$). We interpret this in light of the different assembly histories that lead to different gas fractions at the same halo mass. Gas-poor halos tend to have massive central BHs. The only two mass-accretion channels for simulated BHs are either through multiple mergers (more than average) or a steady gas accretion, both returning thermal energy to their surroundings proportionally to the BH mass excess. However, the distribution of mergers does not support the former scenario. We argue that gas-poor halos have massive BHs because they depleted their gas content due to the AGN feeding/feedback mechanism and star formation, while no gas inflow has replenished their reservoir. However, gas-rich halos have on average lower BH masses since their rapid recent halo and gas accretion have not yet been followed by the BH activity (see also Fig.~\ref{fig:MdotBH_MBH_z_13-13.4}). \cite{booth_dark_2010, booth_towards_2011} found that the BH mass depends on the DM potential and evolution, rather than the properties of the stellar component. It is interesting to notice how very little the instantaneous BH accretion rate is affected by the hot gas fraction in the halo. To trigger this power, it is necessary to funnel the gas to the halo centre, as shown in Fig.~\ref{fig:centralL}. 
\par
It remains unclear to what extent the tight connection between central BH mass and hot gas fraction is a direct consequence of the simplified thermal (two-mode) AGN feedback model implemented in Magneticum (see Sect.~\ref{subsec: 2.1} for details). This model prescribes a thermal energy injection into the surrounding gas particles, proportional to the mass accretion rate, while neglecting other forms of feedback, such as non-isotropic outflows and jets. Similar trends are observed in other state-of-the-art cosmological simulations that employ comparable AGN feedback models, such as EAGLE and IllustrisTNG \citep{davies_quenching_2020}. A crucial step toward a more comprehensive understanding of these dependencies will be the inclusion of more sophisticated AGN feedback prescriptions in cosmological simulations \citep[e.g.][]{sala_supermassive_2024, rennehan_obsidian_2024}.

\subsubsection{The stellar properties}
\label{subsec:3.4.1}
In the second and third panels at the top of Fig.~\ref{fig:Fhotgas_misc}, we show stellar properties, such as the stellar mass in the BCG with central halo $\Delta \log M_\mathrm{\star, \, CEN}$ and the mean stellar formation redshift $\Delta \log z_\mathrm{\star, \, CEN}$, and the fossilness $\Delta \log M_{1,2}$. These quantities are defined according to the star particles bound to the central subhalo in each group or cluster. The fossilness is defined as the logarithmic difference between the stellar masses of the BCG and the second most massive satellite in the halo. This definition often substitutes the magnitude gap used in observation to define the difference between the magnitudes of the central galaxy and the satellites. Such a definition shall describe how evolved the galaxy population is in a halo: a recently accreted or merged group has most likely more massive orbiting satellites than a more relaxed system. We point out that this feature is vastly present in fossil/non-fossil groups \citep{jones_nature_2003}. We define the BCG and its stellar halo as the subhalo 0 in the Subfind catalogue for each group. This corresponds to stars bound to the central galaxy and intracluster or intragroup light (generally, called ICL \citealt{montes_faint_2022}). Stars in the ICL are stars bound to the cluster potential, rather than a single galaxy. Other than yielding different dynamical properties to the BCG stars \citep[e.g.][]{marini_velocity_2021}, the ICL is expected to have fallen into the cluster potential after secular stripping of the infalling satellite galaxies in the halo \citep{contini_formation_2014, contini_origin_2021, brown_assembly_2024, savino_hubble_2025}. Hence, we expect larger stellar halos for systems with older formation times, like the faint sample. 
\par
The stellar properties exhibit a strong anti-correlation with the gas fraction, with the most significant trend observed for stellar mass ($\rho_{\{X, Y\}}=-0.5$). Furthermore, Magneticum predicts that gas-poor systems tend to host a more massive and older stellar component at fixed halo mass. In these systems, the central galaxy forms earlier and harbours a more massive BH, following the $M_{\star, \, CEN}-M_{BH}$ relation, which in turn triggers earlier AGN feedback. Indeed, simulations suggest a connection between dark matter concentration and stellar mass in the local Universe \citep{bassini_black_2019, montenegro-taborda_growth_2023}, consistent with observational and theoretical studies \citep{wang_scatter_2013, tojeiro_galaxy_2017, matthee_origin_2017, zehavi_impact_2018, artale_impact_2018, zu_does_2021, oyarzun_galaxy_2024}.
\par
Moreover, a significant fraction of the stellar mass is likely to originate from infalling satellites, supporting the hypothesis that longer timescales for galaxy processing in the group environment enhance the formation of the ICL. Observational studies of fossil groups align with this scenario \citep{pierini_two_2011, dupke_independent_2022}.

\subsubsection{IGrM, ICM, and satellite properties}
The last row of Fig.~\ref{fig:Fhotgas_misc} reports results for the total (mass-weighted) gas metallicity $Z_{500}$, the temperature $k_B T_{500}$, and the richness $N_\mathrm{gal}$. The mass-weighted temperature and metallicity are calculated using all hot gas non-star-forming particles (i.e. $T>5\times 10^{5} K$) within $R_{500}$.
\par
Metals are forged during the stellar evolution (in the cores of stars and SNe explosions) and the long-term effect of the AGN feedback re-distributes them across the halos \citep{biffi_history_2017, dolag_distribution_2017}. Therefore, older galaxy groups and clusters, with a larger stellar mass, have a larger contribution from metals, compared to younger systems still assembling. Such a contribution is significant only when the metallicity is measured up to larger radii, since we can account for all the metals that have been moved outwards due to gas mixing. However, if the metallicity is measured in the core \citep[see discussion in][]{marini_detecting_2024} we observe no significant correlation with the X-ray luminosity (and, thus, gas fraction). Interestingly, the Spearman coefficient indicates that this effect increases with increasing mass. Similarly, within halos of constant mass, the non-zero Spearman coefficient for the temperature suggests the existence of a multiphase gas in gas-rich halos that has not yet been influenced by gravitational or non-gravitational heating processes. In older and X-ray fainter systems, the temperature is on average higher because the remaining gas has thermalised with the large potential well of the group. Conversely, measuring the number of satellites in $R_{500}$ indicates a higher fraction of galaxies surrounding gas-rich halos, and vice-versa \citep[e.g.][]{puddu_gas-poor_2022}. This is consistent with a denser environment and recent mergers.
\par
We point out that due to the substantially different availability of gas, the star formation rate of central galaxies is also expected to change significantly between (X-ray) bright and faint systems. \cite{vladutescu-zopp_radial_2024} investigated this effect for a higher resolution box than our own (i.e. \textit{Box4/hr}), finding star-forming central galaxies to be generally X-ray brighter than quiescent ones. Recent results from the synthetic eROSITA data \citep{oppenheimer_eagle_2020, vladutescu-zopp_radial_2024} and eFEDS data also indicate the presence of such a correlation \citep{comparat_erosita_2022, chadayammuri_testing_2022}, while the eRASS:4 stacking results contradict this picture \citep{zhang_hot_2025}.

\section{Discussion}
Different studies have investigated the dependence of halo properties such as their formation time, accretion rate, spin, shape, velocity dispersion, concentration, and anisotropy \citep{gao_age_2005, gao_assembly_2007, faltenbacher_assembly_2009, borzyszkowski_zomg_2017, ganeshaiah_veena_cosmic_2018, paranjape_halo_2018} on the environment. While this dependence is not unexpected \citep{sheth_environmental_2004, avila-reese_dependence_2005, gao_age_2005, gao_assembly_2007, weinmann_properties_2006}, the specific role of the non-linear cosmic web in this matter is still highly debated and not completely understood. However, our result connecting the preferential location of X-ray bright halos in denser environments aligns with previous findings that highlight the role of large-scale structures in modulating halo gas content in different structures \citep{romano-diaz_zomg_2017, donnan_role_2022, gouin_gas_2022, gouin_soft_2023, davies_jonathan_j_are_2023, ma_neutraluniversemachine_2024, hasan_filaments_2024, zarattini_where_2024}. 
\subsection{Comparison with past work}
A large pioneering work in Magneticum focused on the effect of the central SMBH activity in the halo properties and mass \citep{steinborn_origin_2016, steinborn_cosmological_2018, castro_impact_2021, angelinelli_mapping_2022, angelinelli_redshift_2023} motivated our investigation to go beyond the typical scenario of AGN feedback effect as uniquely responsible for explaining the X-ray selection effect in galaxy groups. \cite{castro_impact_2021} found that AGN feedback follows a nearly universal time-dependent pattern. The peak of AGN activity occurs slightly before the peak in the baryon fraction, followed by a rapid decline around $z\sim 1$ and a more gradual decrease at lower redshifts. They also observed a strong negative correlation between AGN feedback intensity and halo mass growth, particularly when halos had accumulated 30–50\% of their final mass. This suggests that AGN feedback significantly influences mass loss in an early phase of halo assembly when the gravitational potential well is still relatively shallow, making gas more susceptible to ejection. \cite{ragagnin_simulation_2022} further investigated the AGN-driven gas ejection and its dependence on formation redshift. They found that clusters that formed earlier ($z>1$) tend to have lower gas fractions than those forming at later times, attributing this to the greater efficiency of AGN feedback in expelling gas at earlier epochs when the available gas reservoir was larger. In this same direction, \cite{angelinelli_mapping_2022, angelinelli_redshift_2023} investigated the evolution of the baryonic fraction and its connection with the AGN presence across different epochs, finding it consistent with the emerging picture coming from other state-of-the-art simulations, such as IllustrisTNG, EAGLE, and SIMBA \citep{davies_gas_2019, sorini_how_2022, robson_redshift_2023, ayromlou_feedback_2023}. The consensus is that the AGN feedback is the main engine for shaping the baryonic fraction in galaxy groups and clusters; however, here we argue that this is only one side of the story and different assembly histories -- possibly connected to the larger-scale environment-- must play an important part in driving the scatter of the self-similar evolution. 
\par
\begin{figure*}
  \centering
  \subfloat{\includegraphics[width=\linewidth,trim={0 3.5cm 0 0},clip]{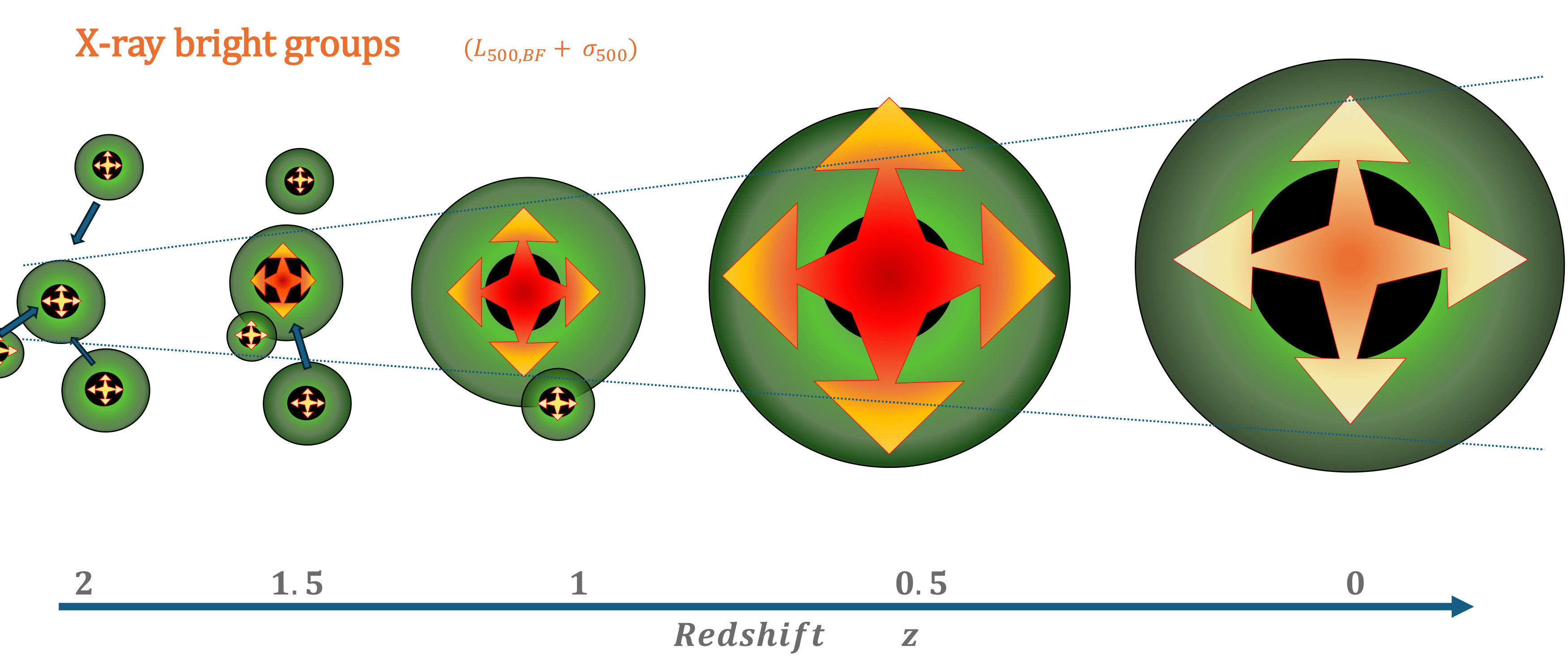} } \\
   \subfloat{\includegraphics[width=\linewidth,trim={0 0 0 0.2cm},clip]{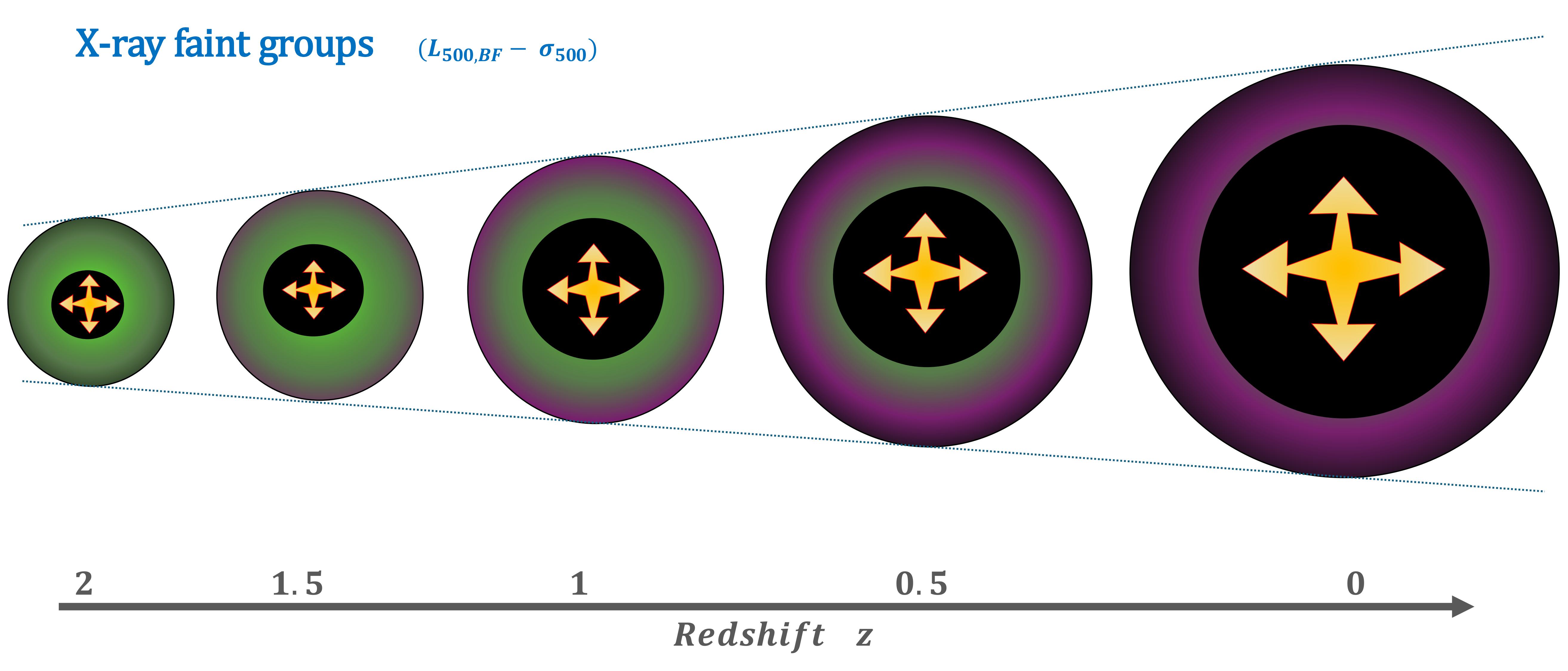}} \\
  \subfloat{\includegraphics[width=\linewidth,trim={0 15.cm 15cm 0},clip]{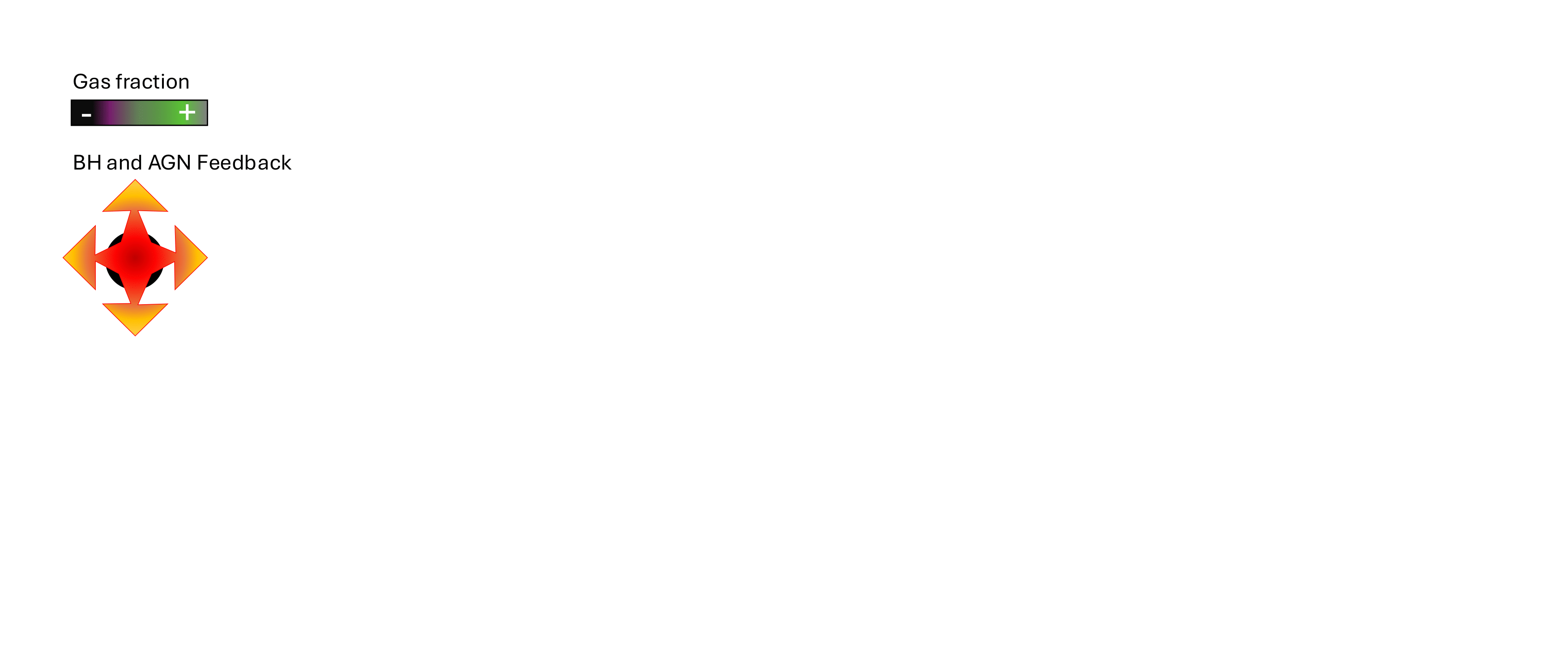} } \\
    \caption{Schematic representation of the assembly history of the bright (top) and faint (bottom) galaxy groups in the simulation. The colour scheme (shown as a green-purple gradient) represents the hot gas fraction in the halo within the virial radius, while the sizes of the circles are proportional to the halo mass. At their core, groups host SMBHs (black circles) triggered by AGN episodes at different intensities (as depicted by the coloured arrows), depending on the surrounding gas content. The BH sizes are exaggerated to highlight the difference in the $M_{BH}/M_\mathrm{halo}$ ratio between the bright and faint samples. The colour and size of the arrow hint at the strength of the AGN activity: redder and larger arrows correspond to a stronger effect. The bright sample is characterised by a fast mass accretion at the late stages of their evolution, given that they are located in dense environments, while the faint one comprises halos evolving relatively isolated in under-dense regions. The dotted lines illustrate the typical self-similar evolution expected for a DM halo with an NFW profile, fitting the evolution of the faint sample.  } \label{fig:bh_accretion}
\end{figure*}

 Figure~\ref{fig:bh_accretion} schematically represents how this accretion history impacts the gas fraction (shown as a green-purple gradient) and the SMBH accretion rate (represented by the size of the central black circle). In response to the newly accreted gas reservoir, AGN feedback is triggered, with its intensity indicated by the colour-coded crossed arrows. In these halos, late-time AGN feedback does not deplete the gas but instead raises the IGrM temperature, contributing to enhanced X-ray emission. In contrast, the faint sample consists of systems that formed with higher masses at earlier epochs but experienced a more isolated evolutionary path, making them likely candidates as ``fossil groups''. As a result, their gas reservoirs have been largely exhausted by $z=0$, leading to negligible X-ray emission. These halos experience an increase in halo mass with $z$ consistent with the changing background cosmic density in the Universe (i.e. dotted lines in the image). This effect is driven by the differences in the hosting environment (see Fig.~\ref{fig:environment_z0}). 
\par
At a given halo mass, galaxy groups at low redshift can exhibit vastly different assembly histories, which in turn shape both the properties of the halo and its host galaxy population. Simulations reveal that X-ray bright groups are characterised by sustained mass accretion at late times ($z < 1.5$), a consequence of the dense environments in which they reside. \cite{davies_jonathan_j_galaxy_2022, davies_jonathan_j_are_2023} reached similar conclusions on the role that mergers play in secular AGN activity and the quenching mechanism for star formation, providing clues on how relevant the difference in implementations of the AGN feedback schemes in different simulations can be \citep{davies_quenching_2020}. Similarly, \cite{cui_hyenas_2024} found that gas fraction is strongly dependent on the halo formation time for groups selected in the HYENAS project.

\subsection{A look out into the future}
We acknowledge that the picture remains complex, with many details still requiring clarification. For instance, some studies report gas-rich halos with high DM concentrations \citep[e.g.][]{andreon_why_2019, popesso_x-ray_2024}. \citet{popesso_x-ray_2024} analysed stacked eFEDS event files for optically selected galaxy groups to derive average surface brightness profiles and compare them to detected systems. Using MAMPOSST \citep{mamon_mamposst_2013}, they reconstructed the underlying DM concentration by modelling the cluster mass and velocity anisotropy profiles from the observed phase-space distribution of member galaxies. \par
However, assigning galaxies to groups can introduce significant biases in velocity dispersion measurements \citep[e.g.][]{old_galaxy_2015}, potentially affecting mass estimates. Optical methods are also highly susceptible to contamination when reconstructing total group and cluster masses. \citet{marini_detecting_2025} highlighted the scatter in halo mass estimates when using velocity dispersion-based definitions. It is in our plan to revisit the stacking procedure outlined in \citet{popesso_x-ray_2024}, taking these effects into account. 

\par
To address these challenges, future work will refine the stacking procedure and DM concentration reconstruction to mitigate such biases. Our primary goal is to assess whether environment-driven accretion and halo assembly bias significantly shape the baryonic content of galaxy groups and clusters, beyond the effects of AGN feedback. When these effects are properly accounted for, stacking can be a powerful tool, even enabling the recovery of spectral-reconstructed properties such as halo temperature \citep{toptun_erosita_2025}.  A step further can be taken by reconstructing filaments and voids in the large scale environment and investigating these properties as a function of the distance from the filaments with tools such as DisPerSE \citep{sousbie_persistent_2011, sousbie_persistent_2011-1} and the Monte Carlo Physarum Machine \citep[MCPM;][]{hasan_filaments_2024}, to name a few. However, this goes beyond the scope of this paper, and we leave this application for future work.
\par 
Additionally, since X-ray emission traces hot gas and is highly sensitive to density, it primarily reflects the most concentrated regions of the IGrM. However, the SZ effect \citep{zeldovich_interaction_1969, sunyaev_observations_1972}, which depends on the integrated pressure, can provide complementary insights -- especially in cases where AGN feedback or mergers redistribute the gas without fully expelling it. This means that while X-ray emission may drop due to gas becoming more diffuse over time, the SZ signal could persist, making it valuable for capturing the full picture of a halo’s baryon content and thermodynamic history \citep[see][for a review]{mroczkowski_astrophysics_2019}. Hydrodynamical cosmological simulations suggest that the integrated SZ signal $Y_\mathrm{SZ}$ is one of the most reliable proxies for total cluster mass, exhibiting a scatter of approximately 10\% \citep[e.g.][]{motl_integrated_2005}. This has motivated the development of low-scatter X-ray mass proxies, such as the core-excised quantity $Y_{X}\equiv M_\mathrm{gas}T_{X}$, introduced by \citet{kravtsov_new_2006}. Given this, jointly analysing X-ray and SZ signals offers a more complete understanding of how feedback and accretion shape the gaseous haloes of galaxy groups and clusters. Future analysis will explore the impact of the SZ effect on recovering the baryonic content at the mass scales.

\section{Conclusions}
In this work, we investigated the key physical processes driving biases in X-ray-selected samples of galaxy groups within the Magneticum simulations. We focused on the role of environment, AGN feedback, and halo assembly history in shaping their gas content, which in turn affects their X-ray properties. Our study is motivated by the vast amount of data eROSITA currently provides at these mass scales \citep{merloni_erosita_2012}.
\par
Galaxy groups exhibit a large scatter in the $L_{500}-M_{500}$ relation, causing a large impact on their X-ray detectability in surveys like eROSITA. The hot gas distribution in galaxy groups is key to their X-ray detectability (described in Fig~\ref{fig:centralL}--\ref{fig:centralS}). Bright systems retain more hot gas than faint ones of the same total mass, but their gas distribution varies:
\begin{itemize}
    \item Centrally luminous groups experience AGN feedback that prevents cooling, increasing gas density and X-ray brightness.
    \item Extended luminous groups have high surface brightness without a strong central peak, but retain more hot gas than faint systems.
\end{itemize}
Ultimately, X-ray bright halos maintain higher gas fractions up to $R_{500}$, while faint systems are baryon-depleted.
\par
Our analysis reveals that the AGN activity primarily depletes haloes of baryons, particularly at early times ($z\sim2-3$), inducing lower X-ray luminosities, but late-time accretion events, such as mergers at $z<1$, can replenish the gas content, increasing the total gas fraction (see Fig.~\ref{fig:growth_z_13-13.4}). This newly acquired gas fuels the central BH, leading to further AGN feedback that raises the plasma temperature and enhances X-ray emissivity, though without immediate baryon depletion (see the schematic representation in Fig~\ref{fig:bh_accretion}). Whether a halo remains gas-rich or depleted depends on its merger history: halos without late-time mergers experience continued baryon loss due to AGN feedback, leading to declining X-ray emissivity, while those undergoing gas-rich mergers can temporarily boost their X-ray brightness. Over sufficiently long timescales, even today’s bright systems may eventually be depleted by AGN activity, aligning with the faint population. This highlights the need to consider both AGN feedback and merger history (associated with the local environment, as illustrated in Fig.~\ref{fig:environment_z0}) in explaining the presence or absence of hot gas in halos, rather than attributing it solely to AGN feedback.
\par
The impact of assembly history extends to other key observables (see Fig.~\ref{fig:Fhotgas_misc}). A supporting piece of evidence is the strong correlation between the hot gas fraction and the central AGN activity. We find that gas-poor halos host more massive central BHs, likely due to sustained gas accretion and feedback depleting the available gas at early times. In contrast, gas-rich halos have, on average, lower BH masses, as their recent halo growth has not yet resulted in significant BH accretion from infalling gas. Such a tight correlation may result as a consequence of the past DM  gravitational potential well \citep{booth_dark_2010, booth_towards_2011}. 
\par
We confirm that gas-poor halos tend to host more massive and older stellar populations (illustrated in Fig.~\ref{fig:Fhotgas_misc}), consistent with early-formed systems having a longer time for accretion and star formation. Similarly, we observe a strong anti-correlation between the hot gas fraction and the system’s fossilness, suggesting that groups with a more dominant central galaxy have undergone significant early gas depletion. Gas-poor systems exhibit enhanced metal content due to longer stellar evolution and AGN-driven redistribution of metals. AGN activity also affects the temperature, with continuous thermal feedback at late times. Additionally, we find that gas-poor halos tend to reside in less dense environments, as indicated by a lower number of satellite galaxies within $R_{500}$. Finally, our results suggest that the star formation activity of central galaxies differs significantly between X-ray bright and faint systems, as a natural consequence of the gas depletion in X-ray faint groups. 
\par
Overall, our study stresses the complex interplay between AGN feedback, gas accretion, and halo assembly in determining the observable properties of galaxy groups. Future work should further investigate these trends using higher-resolution simulations, observational comparisons, and refined models for BH growth and feedback. Current and upcoming X-ray surveys, such as those from eROSITA \citep{merloni_erosita_2012}, NewAthena \citep{cruise_newathena_2025}, and possibly AXIS \citep{reynolds_overview_2023}, combined with SZ facilities such as the Simons Observatory \citep{abitbol_simons_2025} and AtLAST \citep{di_mascolo_atacama_2024} will provide valuable data to test these predictions and enhance our understanding of galaxy group evolution.

\begin{acknowledgements}
      The authors thank Andrea Biviano and Magda Arnaboldi for their useful discussions. This project has received funding from the European Research Council (ERC) under the European Union’s Horizon Europe research and innovation programme ERC CoG (Grant agreement No. 101045437). KD acknowledges support by the COMPLEX project from the European Research Council (ERC) under the European Union’s Horizon 2020 research and innovation program grant agreement ERC-2019-AdG 882679. The calculations for the Magneticum simulations were carried out at the Leibniz Supercomputer Center (LRZ) under the project pr83li. SVZ acknowledges support by the \emph{Deut\-sche For\-schungs\-ge\-mein\-schaft, DFG\/} project nr. 415510302. NM acknowledges funding by the European Union through a Marie Sk{\l}odowska-Curie Action Postdoctoral Fellowship (Grant Agreement: 101061448, project: MEMORY). Views and opinions expressed are however those of the author only and do not necessarily reflect those of the European Union or of the Research Executive Agency. Neither the European Union nor the granting authority can be held responsible for them. GP acknowledges financial support from the European Research Council (ERC) under the European Union’s Horizon 2020 research and innovation program HotMilk (grant agreement No. 865637) and support from the Framework per l’Attrazione e il Rafforzamento delle Eccellenze (FARE) per la ricerca in Italia (R20L5S39T9). 
\end{acknowledgements}

%
   \bibliographystyle{aa} 
   \bibliography{lib} 
%

\begin{appendix}
\label{appendix}
    \section{Additional figures}
    We provide here the scatter plot between the luminosity in the soft band of X-ray ($0.5-2.0$ keV) and halo mass for the same groups in Fig.~\ref{fig:Fhotgas_misc}. The Spearman correlation has the same trends as with the hot gas fractions.
    \begin{figure}[htbp]
    \centering   
    \subfloat{\includegraphics[width=0.3475\textwidth]{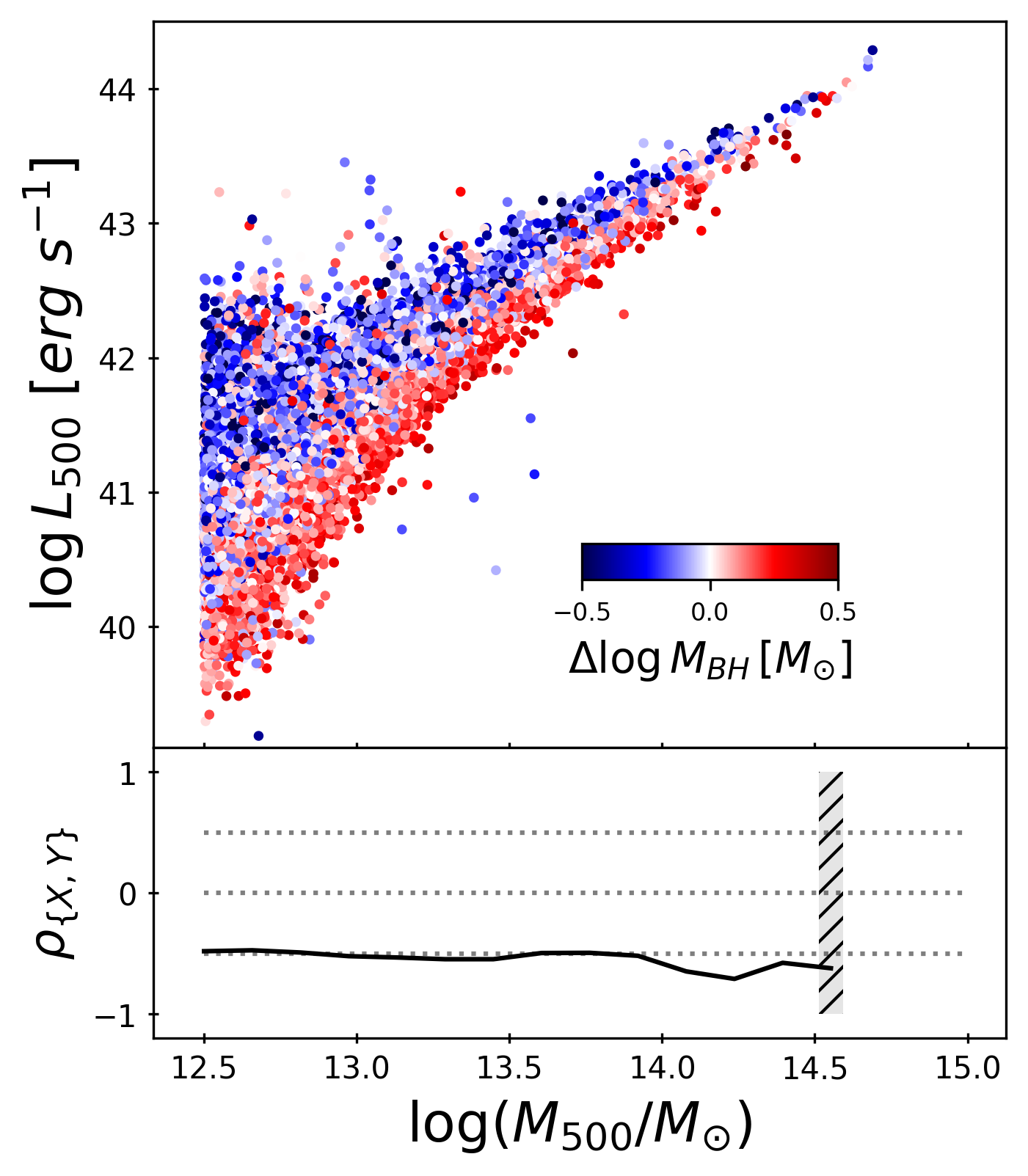}}  
    \subfloat{\includegraphics[trim={1.6cm 0 0 0}, clip, width=0.3\textwidth]{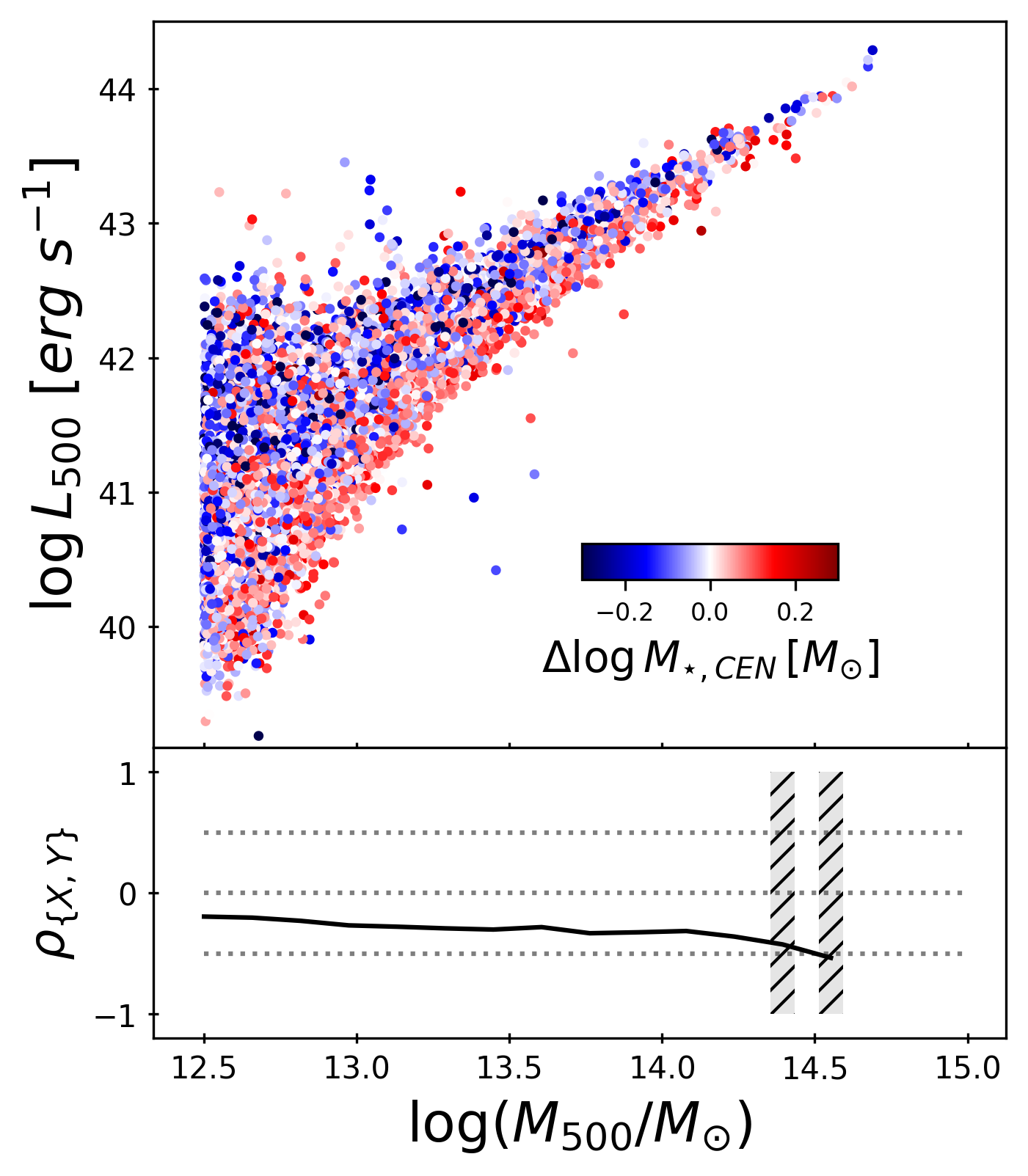}}  
    \subfloat{\includegraphics[trim={1.6cm 0 0 0}, clip, width=0.3\textwidth]{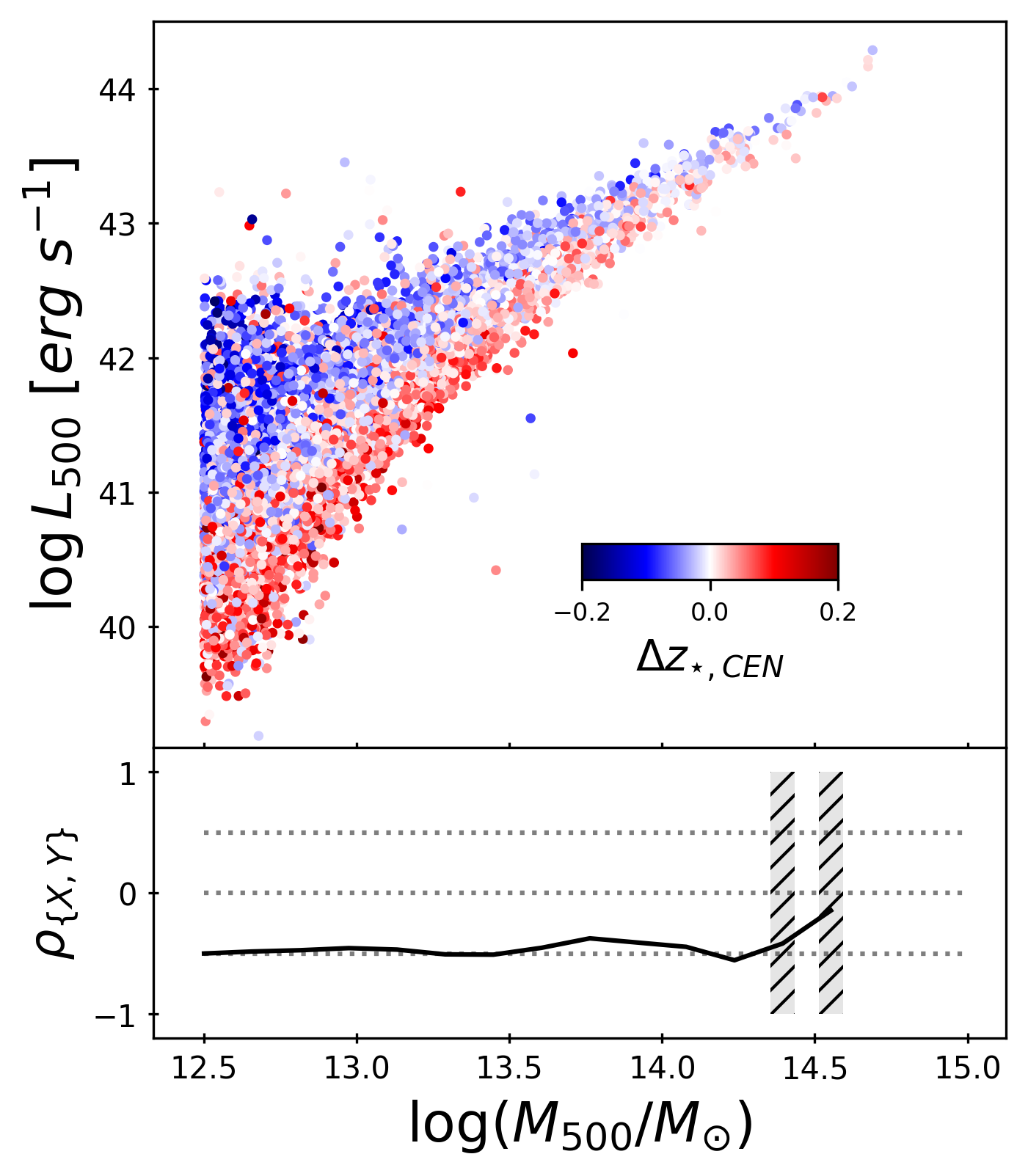}} \\
    \subfloat{\includegraphics[width=0.27\textwidth]{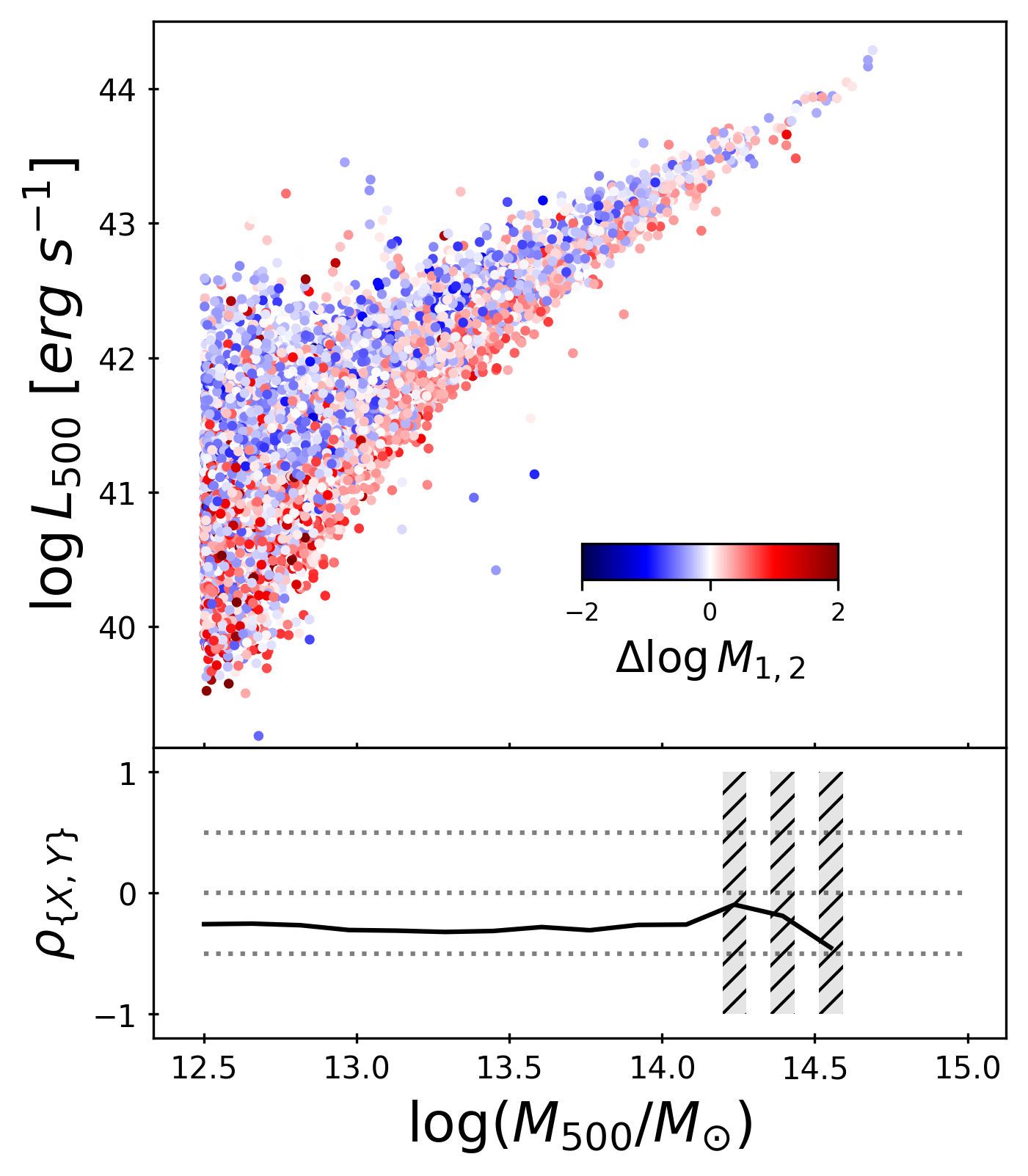}}  
    \subfloat{\includegraphics[trim={1.6cm 0 0 0}, clip, width=0.235\textwidth]{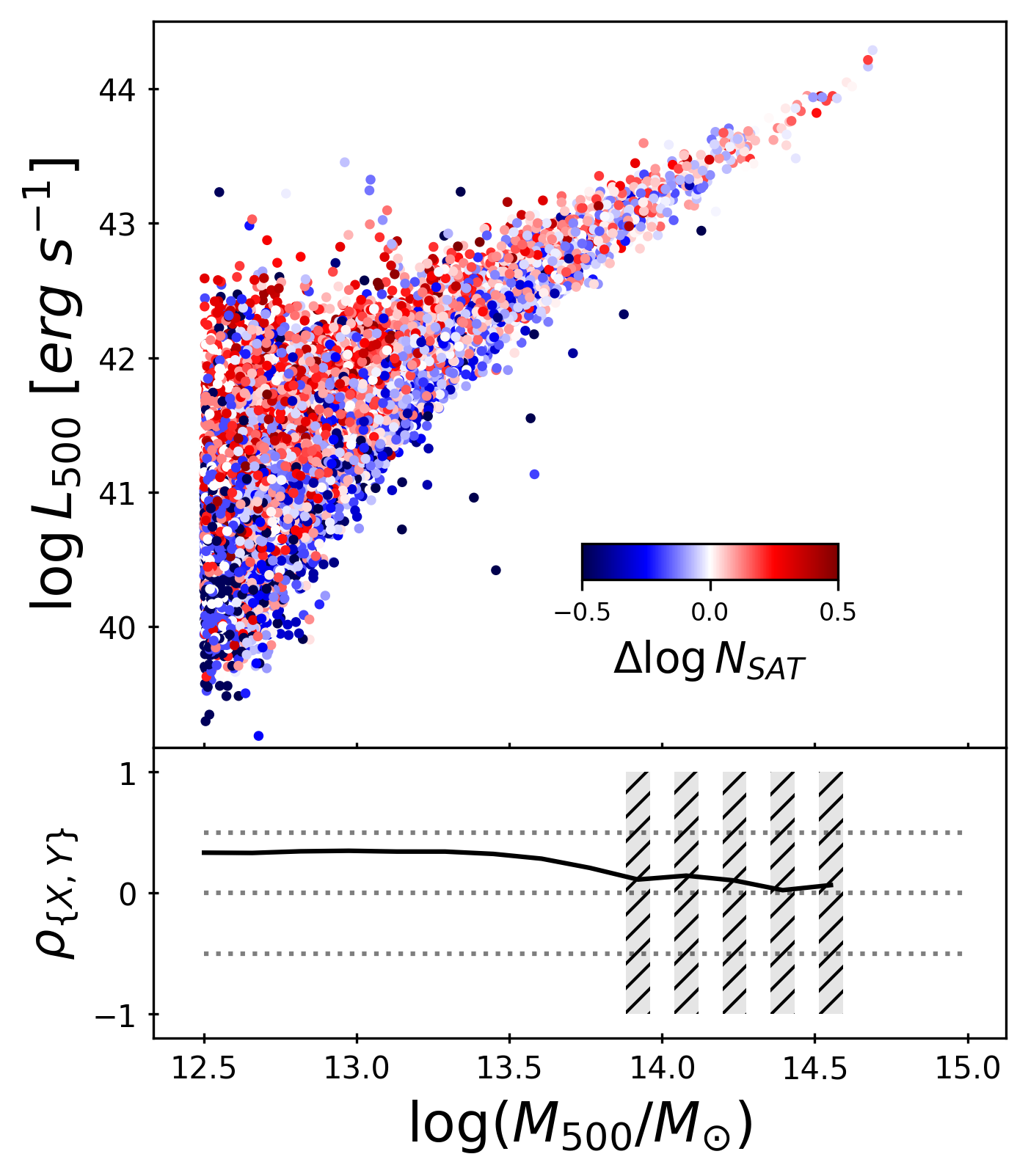}} 
    \subfloat{\includegraphics[trim={1.6cm 0 0 0}, clip, width=0.235\textwidth]{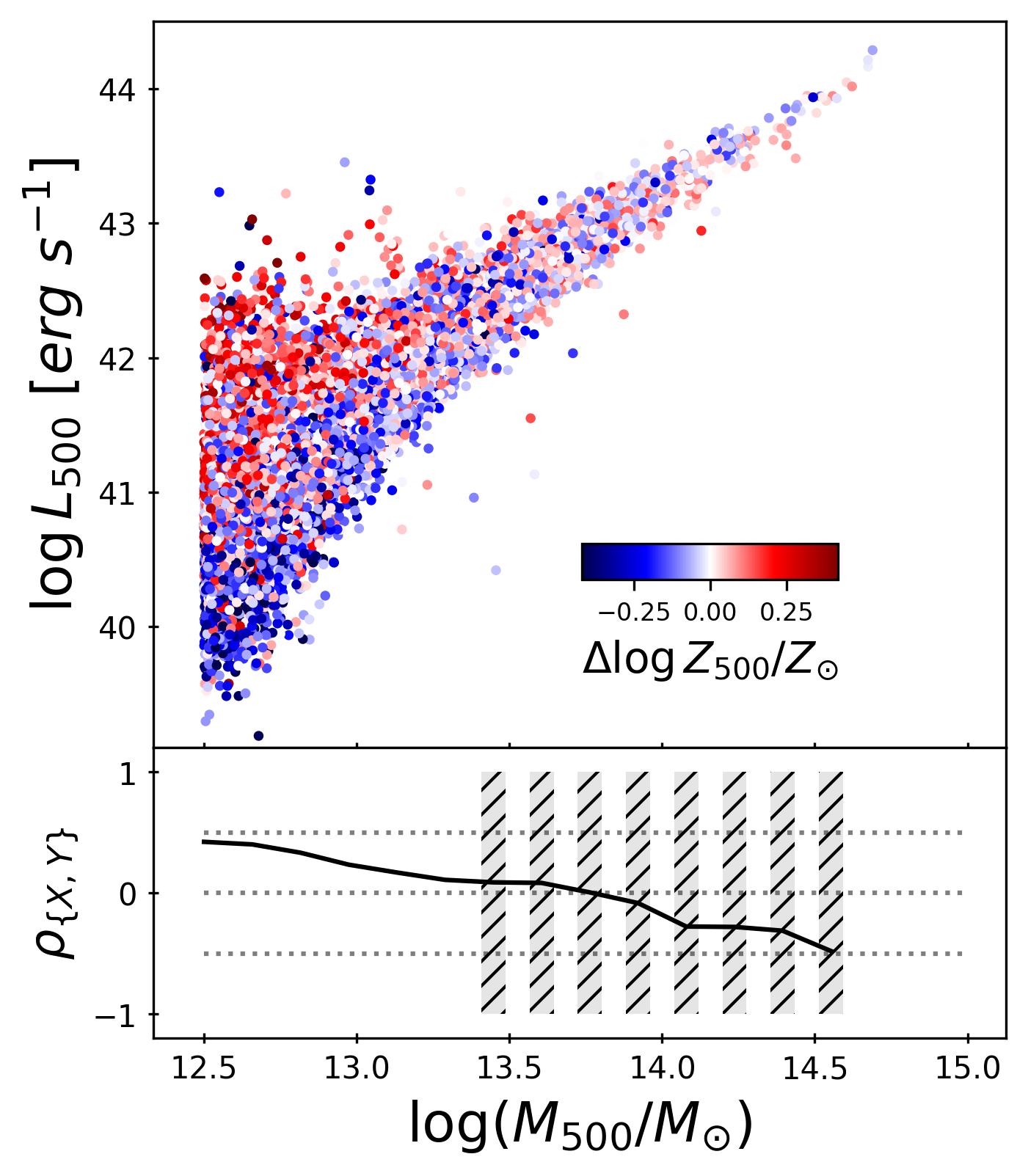}} 
    \subfloat{\includegraphics[trim={1.6cm 0 0 0}, clip, width=0.235\textwidth]{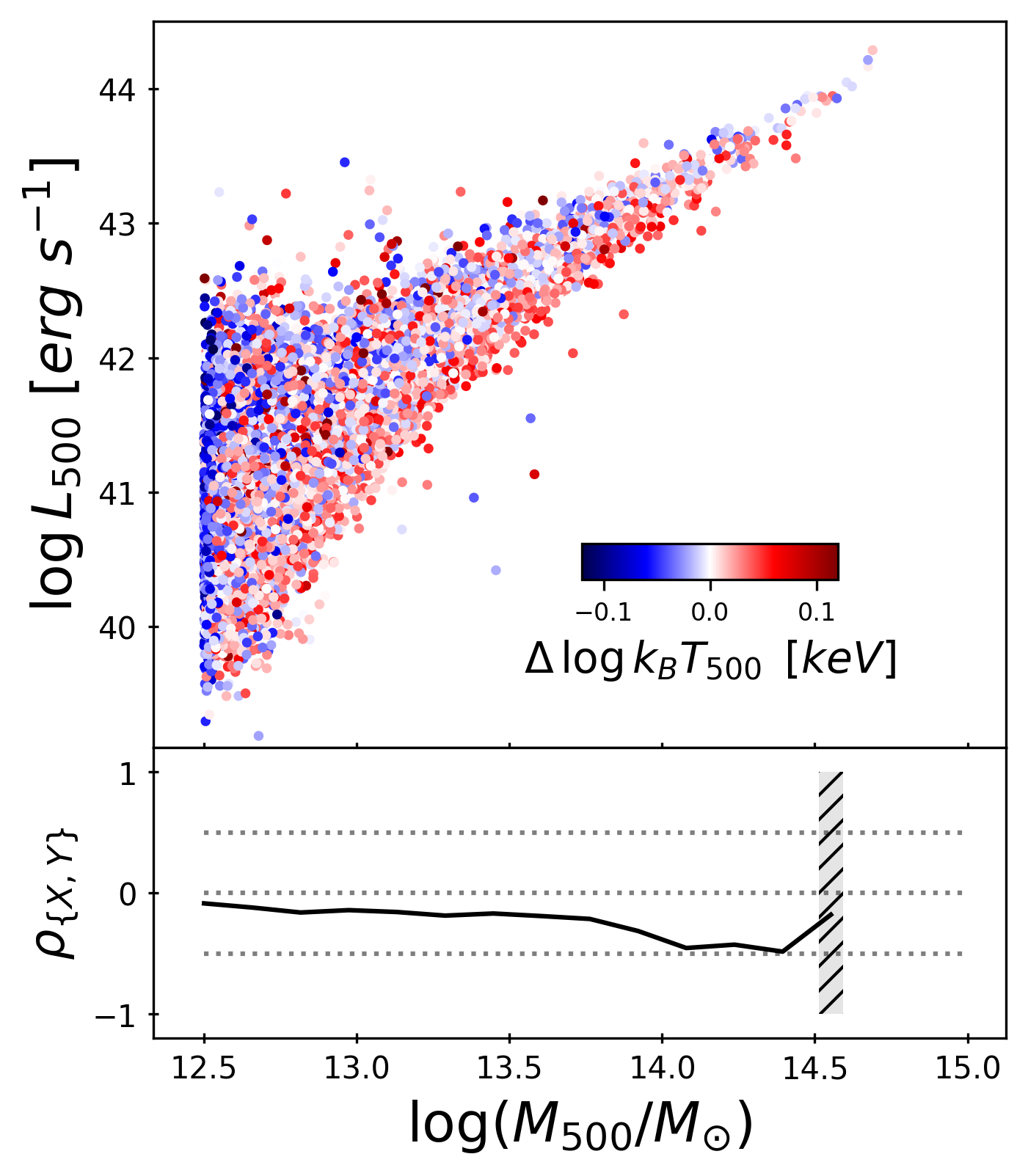}}\\

        \caption{The same plot as in Fig.~\ref{fig:Fhotgas_misc}, using the X-ray luminosity ($0.5-2.0$ keV) as opposed to the hot gas fraction. }
    \label{fig:appendix}
    \end{figure}

\end{appendix}

\end{document}